\documentclass{iau} 
\usepackage{graphicx}
\usepackage{natbib}
\usepackage{aas_macros}
\usepackage{amsmath}
\def\Rs{R_{\odot}}
\newcommand{\bm}{\mathbf}
\newcommand{\pd}{\partial}

\title[IAUS 362~~Predictive Power of Computational Astrophysics as a Discovery Tool] %% give here short title %%
{Advances and Challenges in Observations and Modeling of the Global-Sun Dynamics and Dynamo}

\author[A.G. Kosovichev, G. Guerrero, A.M. Stejko, V.V. Pipin \& A.V. Getling]   %% give here short author list %%
{Alexander Kosovichev$^{1,2}$ Gustavo Guerrero$^{1,3}$, Andrey Stejko$^1$\\ Valery Pipin$^4$ \and Alexander Getling$^5$}

\affiliation{$^1$Center for Computational Heliophysics, New Jersey Institute of Technology, \\ University Heights, Newark, NJ 07102, U.S.A.\\
	email: {\tt alexander.g.kosovichev@njit.edu} \\
$^2$ NASA Ames Research Center, Moffett Field, CA 94035 U.S.A.\\
$^3$Physics Department, Universidade Federal de Minas Gerais\\
Av. Antonio Carlos, 6627, Belo Horizonte, MG 31270-901, Brazil\\
email: {\tt guerrero@fisica.ufmg.br}\\
$^4$Institute of Solar-Terrestrial Physics, Russian Academy of Sciences\\ 
Irkutsk, 664033, Russia\\
email: {\tt pip@iszf.irk.ru}\\
$^5$Skobeltsyn Institute of Nuclear Physics, Lomonosov Moscow State
University,\\ Moscow, 119991 Russia\\
email: {\tt A.Getling@mail.ru}
}

\pubyear{2022}
\volume{362}  %% insert here IAU Symposium No.
\jname{Predictive Power of Computational Astrophysics as a Discovery Tool}
\editors{A.C. Editor, B.D. Editor \& C.E. Editor, eds.}
\begin{document}

\maketitle

\begin{abstract}
%Numerical simulations play a crucial role in understanding the complex multi-scale turbulent processes in the solar interior. 
Computational heliophysics has shed light on the fundamental physical processes inside the Sun, such as the differential rotation, meridional circulation, and dynamo-generation of magnetic fields. However, despite the substantial advances, the current results of 3D MHD  simulations are still far from reproducing helioseismic inferences and surface observations. The reason is the multi-scale nature of the solar dynamics, covering a vast range of scales, which cannot be solved with the current computational resources. In such a situation, significant progress has been achieved by the mean-field approach, based on the separation of small-scale turbulence and large-scale dynamics. The mean-field simulations can reproduce solar observations, qualitatively and quantitatively, and uncover new phenomena. However, they do not reveal the complex physics of large-scale convection, solar magnetic cycles, and the magnetic self-organization that causes sunspots and solar eruptions. Thus, developing a synergy of these approaches seems to be a necessary but very challenging task.
\keywords{Sun: activity, Sun: helioseismology, Sun: magnetic fields, Sun: interior, MHD, turbulence}
\end{abstract}

\firstsection % if your document starts with a section,
\section{Introduction}
The 11-year sunspots cycles have been regularly observed for more than 400 years. Tremendous observational material has been accumulated, but a basic understanding of the underlying physical processes is still lacking. There is no doubt that the magnetic cycles are caused by dynamo processes in the deep solar interior. These processes are associated with the interaction of turbulent convection with solar rotation beneath the solar surface, understanding of which is challenging. The observational data provided by helioseismology have limited spatial and temporal resolutions and provide only information about large-scale flows. The internal magnetic field has not been determined directly. Helioseismic inferences suffer from substantial systematic uncertainties because of so-called realization noise due to the random excitation of solar oscillation and complicated interaction between oscillations and convection. Helioseismic techniques and results must be tested by using numerical models and simulations of solar oscillations. A satisfactory theoretical understanding can be achieved only through realistic modeling and numerical simulations.

The realistic computational modeling of the solar interior and atmosphere requires resolving a wide range of scales from small-scale flows of granulation to global-scale zonal and meridional flows. The flows are characterized by extreme Reynolds numbers so that the full range of dynamical scales can never be resolved. However, high-resolution simulations of the near-surface convection have shown that the observed phenomena can be modeled with satisfactory accuracy using subgrid-scale turbulence models, such as the Large-Eddy Simulation (LES) methods. This method aims to resolve all dynamically essential scales and approximate the dissipation scales by applying turbulence scaling laws. The LES method requires that the simulations are performed with a grid resolution of 25-100 km. Therefore, the computational models are limited to small `box' regions covering $\lesssim 100$ Mm horizontally and even less in depth \citep[e.g.][]{Rempel2009,Stein2012,Kitiashvili2015}. Nevertheless, these simulations reproduce the solar subsurface dynamics with a high degree of realism and have their prediction power. In particular, they predicted the formation of small-scale vortex structures in the solar granulation, uncovered mechanisms of acoustic wave excitation by turbulent convection, explained magnetic self-organization of solar magnetic fields, plasma flows in sunspots (the Evershed effects), local MHD dynamo, etc.

The global-Sun modeling is currently performed using two basic approaches: 1) by solving the 3D MHD equation in an anelastic approximation \citep{Gough1969} for suppressing the fast wave motions and excluding the surface and near-surface layers, typically, at $r > 0.95 R_\odot$; 2) by separating turbulent and large-scale flows by applying a mean-field approximation.  

The 3D MHD computational models provide great insight into the global Sun dynamics and dynamo mechanisms \citep[e.g.][]{Ghizaru2010,Charbonneau2013,Miesch2011,Simitev2015,Guerrero2019}. However, they have been unable to reproduce the observations. The reason is that currently, the 3D MHD models cannot resolve the turbulent convective motion with the resolution necessary for describing the non-linear interaction of these motions with rotation and magnetic field -- such interaction results in highly anisotropic turbulent transport affecting the large-scale momentum and energy balance. 

A theoretical description found in the mean-field approximation separates the small-scale turbulent motions from large-scale flows and describes their interaction in turbulent transport coefficients depending on large-scale properties \citep[e.g.][]{Krause1980,Brandenburg1992,Kitchatinov1994,Pipin2000}. In addition, to turbulent diffusion, this approach predicts non-diffusive turbulent effects, which appear as additional terms in the mean-field equations. This theory has been developed to a high degree of sophistication. However, it cannot predict the strength and spectrum of turbulence and, in solar-stellar applications, uses parameters of convective velocity distribution from the mixing-length anzats of the stellar evolution theory. Nevertheless, the mean-field theory provides a reasonable qualitative and quantitative description of the differential rotation, meridional circulation, solar-cycle evolution of the axisymmetric magnetic field and makes predictions that can be verified in observations. Traditionally, the dynamical and dynamo processes were considered separately. But, recently developed combined full-MHD mean-field models explained many observed phenomena, such as the differential rotation, migrating zonal flows (`torsional oscillations'), extended solar cycle phenomenon for zonal and meridional flows \citep{Pipin2019,Pipin2020}. Despite the success, this theory cannot describe the observed 3D structure of the solar convection and magnetic fields. Although, initial steps towards a 3D mean-field description have been made.

This paper presents recent attempts to develop a synergy of the global 3D MHD modeling, mean-field theory, and helioseismic analysis by our group. By no means is this a comprehensive review of the current advances to develop computational models of the global-Sun dynamics and magnetism, to which many other groups have contributed.

\section{Helioseismic Observations of the Solar Dynamo and Interior Dynamics}

\begin{figure}[t]
	% \vspace*{-2.0 cm}
	\begin{center}
		\includegraphics[width=4.0in]{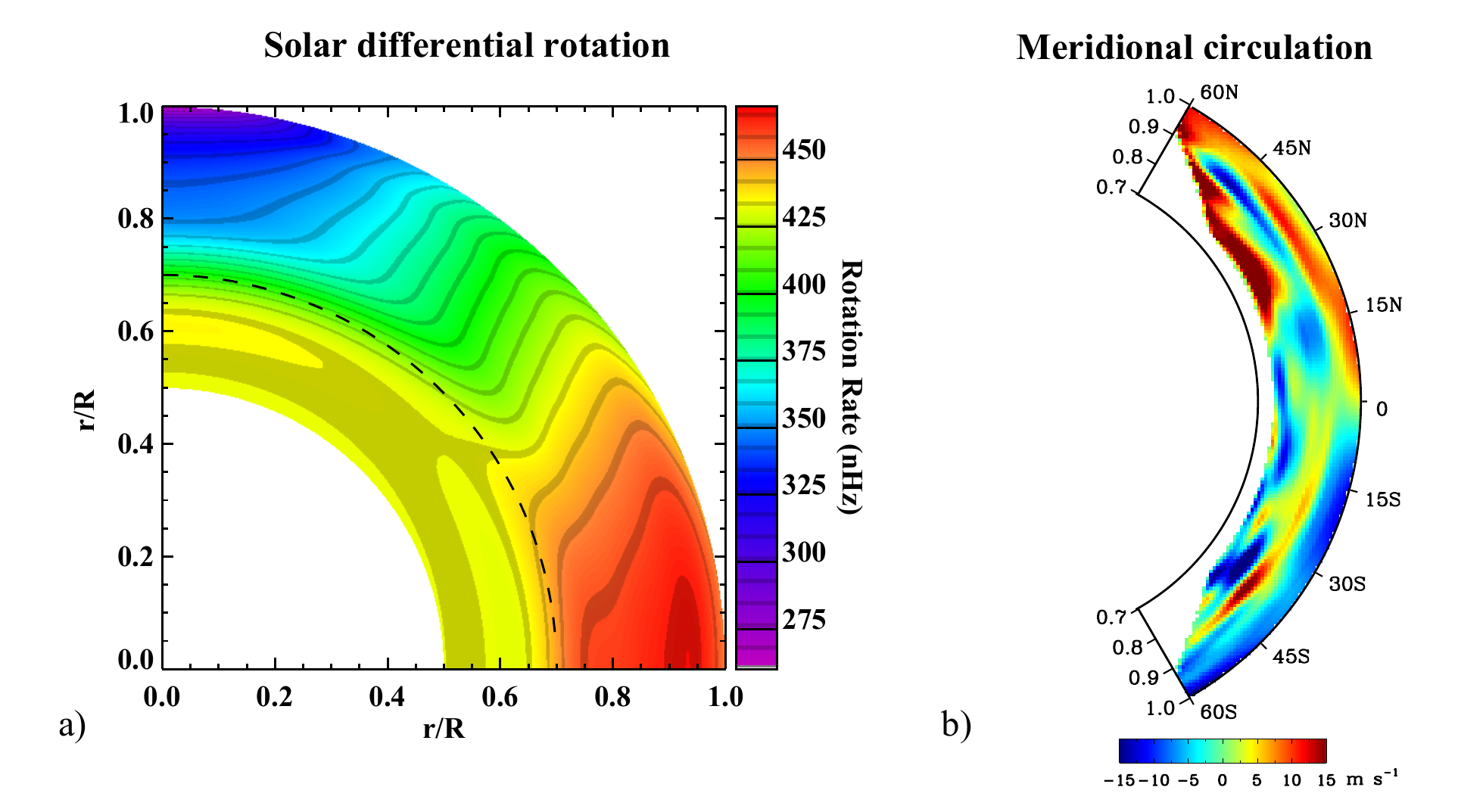} 
		% \vspace*{-1.0 cm}
		\caption{a) Mean solar rotation determined by global helioseismology  from SDO/HMI JSOC data in 2010-2022. b) The distribution of the meridional flow velocity in the solar convection zone \citep{Zhao2013}. 
		}
		\label{fig1}
	\end{center}
\end{figure}

The helioseismic observations are generally divided into two complementary approaches: global and local helioseismology. Global helioseismology measures frequencies and frequency splitting of resonant acoustic oscillations and surface gravity waves, inversion of which provides estimates of the internal rotation (Fig.~\ref{fig1}a) and axisymmetrical sound-speed variations in the whole spherical Sun. The measurements are typically performed using uninterrupted 72-day series of solar 5-minute oscillations. Only azimuthally averaged flows and structures can be measured by this technique. 

Local helioseismology measures either local variations of the waves dispersion relation by using 3D Fourier transform \citep[Ring-Diagram Analysis, ][]{Gough1983,Hill1989} or acoustic travel times extracted from cross-correlations of solar oscillations \citep[Time-Distance Helioseismology, ][]{Duvall1993,Kosovichev1997}. These techniques provide maps of horizontal flows in shallow subsurface layers, albeit with different resolutions. In addition, time-distance helioseismology has been used to measure large-scale meridional flows through the whole of the solar convection zone. 

Helioseismic measurements of the internal differential rotation, meridional circulation, and their variations with the solar cycle are essential for understanding solar dynamics and magnetism. It has been firmly established \citep{Schou1998} that the solar differential rotation extends through the whole convection zone and has strong shearing flows: the tachocline at the bottom and the Near-Surface Shear Layer (NSSL) the top boundaries (Fig.~\ref{fig1}a). The shearing flows play key roles in the solar dynamo processes. The tachocline is a likely place of magnetic field generation, and the NSSL `shapes' the evolution of magnetic fields observed on the surface, in particular, forming the equator-ward migration of emerging magnetic flux - the butterfly diagram. 

The distribution of the meridional circulation with depth in the convection zone is much less certain. The initial measurements \citep[Fig.~\ref{fig1}b, ][]{Zhao2013} showed that the circulation might consist of two circulation cells along the solar radius, with the polar-ward flows at the top and bottom of the convection zone and return equator-ward flow in the middle, contrary to the expectations that the circulation has a simple single-cell structure. This result caused significant attention and new measurement efforts because, if correct, it effectively rules out the flux-transport solar-cycle models, which predict the polar-ward meridional flow is responsible for the butterfly diagram. However, the exact flow structure has not been reliably established. Some studies show that the single-cell structure is more likely, while some others find even more complex multi-cell dynamics \citep{Boning2017,Chen2017,Chen2018,Gizon2020a,Jackiewicz2015a,Kholikov2014b,Lin2018a,Schad2013a}. Additional uncertainty comes from potential variations of the meridional flows with the Sun's activity cycle.Because the measurements require very long time series of solar oscillations, the results may depend on the time interval chosen for the analysis. 

In such a situation, forward 3D acoustic modeling of helioseismic data is essential for resolving the discrepancies and determining the observational limits on the flow dynamics \citep{Hartlep2013,Hartlep2008,Khomenko2009,Parchevsky2009,Parchevsky2014}. Some recent advances in this direction are discussed in the next section.

\section{Computational Solar Acoustics and Forward Modeling of Helioseismic Inferences}

Helioseismic inferences of the internal structures and flows are under simplified assumptions about wave propagation. Relationships between the observed quantities, such as frequency shifts and travel-time anomalies, and the physical properties are obtained using a perturbation theory and expressed in the form of linear integral equations. The inversion procedures for solving these equations seek for the smoothest solution satisfying the relationships within the observational errors. Thus, the real observational constraints on complex flow structures remains unknown.

For getting insight into uncertainties of helioseismic measurements and obtain observational constraints the GALE (Global Acoustic Linearized Euler) code \citep{Stejko2021} had been developed. The code solves the conservation form of the linearized compressible Euler equations on a full global 3-dimensional grid: $0 \le \phi \le 2\pi$, $0 < \theta < \pi$, $0 < r \le R_{\odot}$. 

\begin{equation}\label{eq:gov1}
	\dfrac{\partial \rho'}{\partial t} + \Upsilon' = 0 \ ,
\end{equation}
\begin{equation}\label{eq:gov2}
	\dfrac{\partial\Upsilon'}{\partial t} + \boldsymbol{\nabla}:\left(\mathbf{m}'\tilde{\mathbf{u}} + \tilde{\rho}\tilde{\mathbf{u}}\mathbf{u}'\right) = -\nabla^{2}\left(p'\right) - \nabla\cdot\left(\rho'\tilde{g}_{r}\mathbf{\hat{r}}\right) + \nabla\cdot S\mathbf{\hat{r}} \ ,
\end{equation}
\begin{equation}\label{eq:gov3}
	\dfrac{\partial p'}{\partial t} = - \dfrac{\Gamma_{1}\tilde{p}}{\tilde{\rho}}\left(\nabla\cdot\tilde{\rho}\mathbf{u}'+ \rho'\nabla\cdot\tilde{\mathbf{u}} - 
	\dfrac{p'}{\tilde{p}}\tilde{\mathbf{u}}\cdot\nabla\tilde{\rho} + \tilde{\rho} \mathbf{u}'\cdot\dfrac{N^{2}}{g}\mathbf{\hat{r}}\right) \ .
\end{equation}

Here the acoustic perturbations are denoted by a prime, and the background field terms are denoted by a tilde. $\Upsilon$ is defined as the divergence of the momentum field $\mathbf{m}$ ($\Upsilon = \nabla\cdot\mathbf{m}= \nabla\cdot\rho\mathbf{u}$), computing perturbations in the acoustic (potential) flow field and omitting solenoidal terms in our governing equations. The acoustic oscillations are initiated by a randomized source function ($S$), modeling the stochastic excitation of acoustic modes at the top boundary of the convection zone. Governing equations (\ref{eq:gov1}) - (\ref{eq:gov3}) are solved using a pseudo-spectral computational method through spherical harmonic decomposition ($f = \sum_{lm} a Y_{lm}$) of field terms ($\Upsilon$, $\rho$, $p$, $\mathbf{u}$) using the Libsharp spherical harmonic library \citep{Reinecke2013}. The governing equations (Eqs \ref{eq:gov1} - \ref{eq:gov3}) are solved in a vector spherical harmonic (VSH) basis, while the material derivative is solved in its Cauchy conservation form ($\boldsymbol{\nabla}:\left(\mathbf{m}'\tilde{\mathbf{u}} + \tilde{\rho}\tilde{\mathbf{u}}\mathbf{u}'\right)$) using a tensor spherical harmonic (TSH) basis. This formulation allows for the use of recursion relations to compute derivatives tangent to the surface of the sphere (${\partial}/{\partial\theta},{\partial}/{\partial\phi}$), resulting in a system of 1D equations for radial relations. \\

\begin{figure}[t]
	% \vspace*{-2.0 cm}
	\begin{center}
		\includegraphics[width=0.9\linewidth]{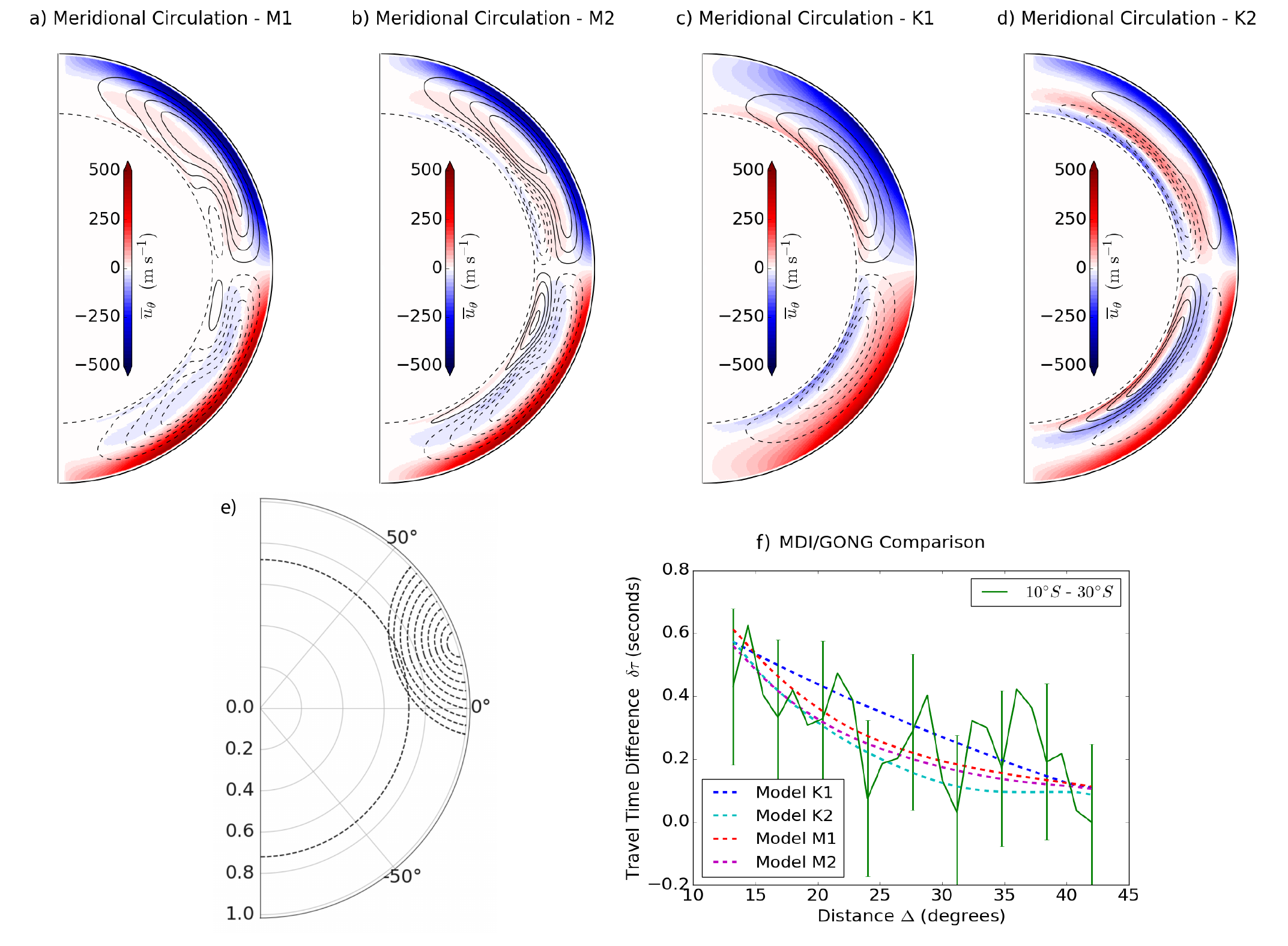} 
		% \vspace*{-1.0 cm}
		\caption{a) The mean-field models of meridional circulation \citep{Pipin2018,Pipin2019}: a) Single-cell meridional circulation with a shallow return flow at $\sim 0.80R_{\odot}$. b) Double-cell meridional circulation with a weak reversal near the bottom of the convection zone. c) Single-cell meridional circulation with a deep return flow near the base of the tachocline. d) Double-cell meridional circulation with a strong reverse flow. Solid and dashed contours represent counterclockwise and clockwise rotation respectively. Meridional circulation speed is increased by a factor of 36 to reduce the duration of computational runs. e) The acoustic ray-paths associated with the selected travel distances for the deep-focus time-distance helioseismology method. f) The North-South travel-time differences ($\delta\tau_{NS}$) as a function of travel distance ($\Delta$) for the models obtained from the helioseismic observations (solid curve) and the computational models (dashed lines).}
		\label{fig2}
	\end{center}
\end{figure}

We use the GALE code to perform forward modeling of helioseismic data and acoustic travel times for various models of the meridional circulation \citep{Stejko2021a}. Then, by comparing the forward-modeling results with the available observational data, we validate the meridional circulation models. In particular, we performed the forward for four meridional circulation models shown in Figure \ref{fig2}a-d. Models M1 and M2 are the self-consistent mean-field dynamo models \citep{Pipin2019} and K1 and K2 are the mean-field hydrodynamics models without magnetic field. The meridional circulation in these models can consist of one or two cells along the radius. The two-cell circulation appears when the radial dependence of the Coriolis number is taken into account \citep{Bekki2017}, which is due to the radial dependence of the convection turn-over time. 
\begin{figure}
	% \vspace*{-2.0 cm}
	\begin{center}
		\includegraphics[width=0.65\linewidth]{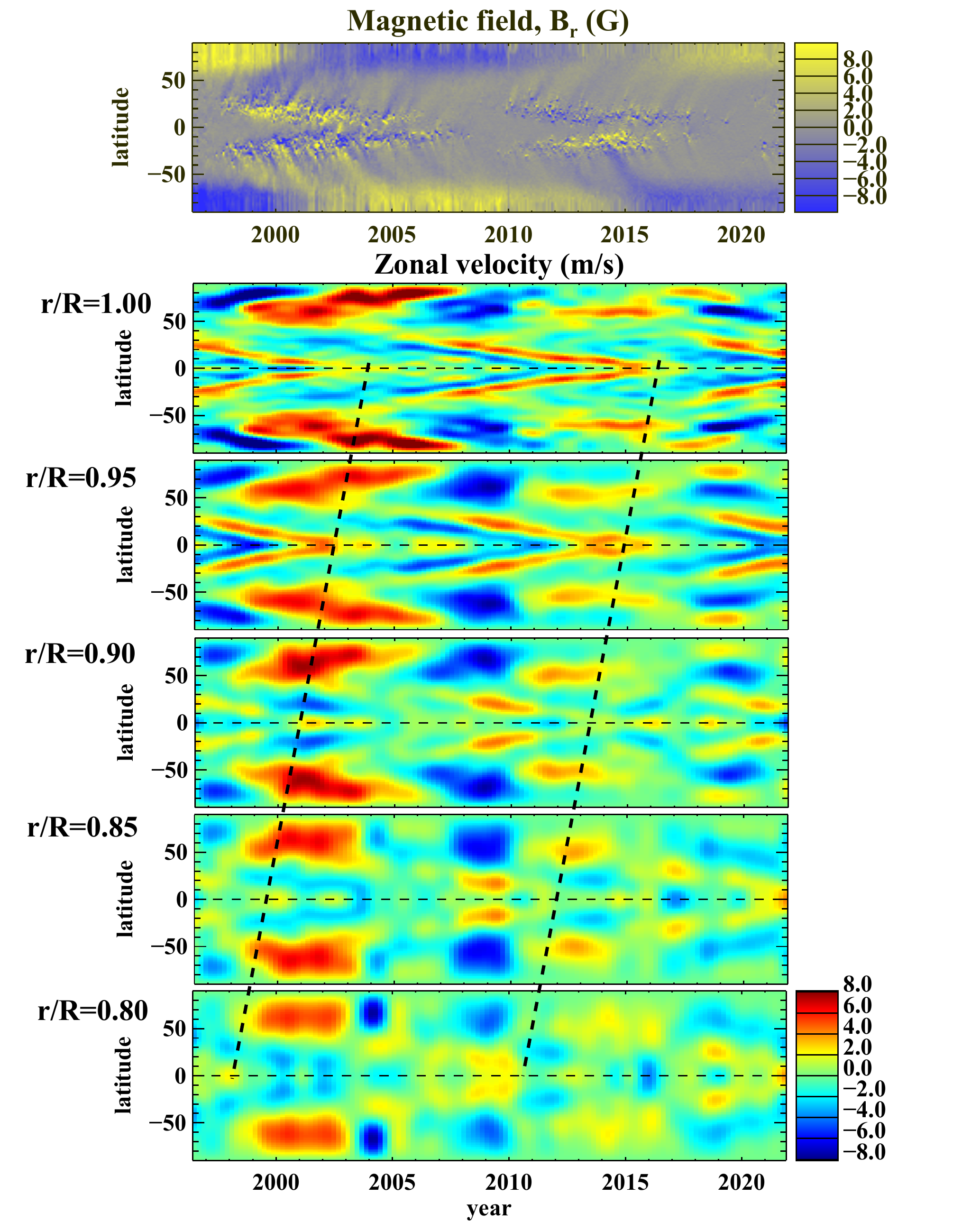} 
		% \vspace*{-1.0 cm}
		\caption{Time-latitude diagrams for the radial magnetic field in 1996-2021, obtained from SoHO/MDI, SDO/HMI and SOLIS data, and the zonal flow velocity calculated from the solar rotation inversion data of SoHO/MDI and SDO/HMI, available from JSOC. The inclined dashed lines track the points of convergence of the fast rotating zonal flows at the equator, indicating the end of the extended Cycles 22 and 23.}
		\label{fig3}
	\end{center}
\end{figure}

The acoustic oscillations generated for these background flow models were analyzed using the same deep-focusing scheme for selecting the acoustic ray paths (Fig.~\ref{fig2}e) and the same cross-covariance fitting procedure, as in the analysis of the solar observations \citep{Zhao2013}. The signal-to-noise ratio for the travel-time measurements was improved by applying specially designed phase-speed filters. The calculated travel times are plotted as a function of the distance between the correlation points of the surface, which corresponds to different depths of the acoustic turning points. The solid curve with error bars shows the measurements obtained from solar observation \citep{Gizon2020}. Evidently, models M1, M2 and K2 with relatively shallow return flows are consistent with the observations, irrespective of whether the meridional circulation has a single- or double-cell structure. But, model K1 with a deep return flow significantly deviates from the observations.

%\newpage
\section{Torsional Oscillations, Extended Solar Cycle and Evidence for Dynamo Waves}

Migrating during the solar cycles zonal flows (`torsional oscillations') carry essential information about the global-Sun dynamics and dynamo processes. The torsional oscillations were first discovered in Doppler-shift measurements of the surface rotational velocity as zones of faster and slower solar rotation. Helioseismic observations showed that the torsional oscillation flows are extended into the deep interior. Figure~\ref{fig3} shows the flow evolution during the last two sunspot cycles in the upper half of the convection zone. It is evident that the flow pattern, which consists of a prominent polar-ward branch and a weaker equatorward branch repeats every 11-years (like the sunspot cycle), but the duration of this pattern is about 22 years. This phenomenon is called the `extended solar cycle'. Both polar and equatorial  branches of faster rotation start at about 60 degrees latitude, approximately at the start of sunspot cycles. The polar branches relatively quickly migrate to the polar regions while the equator-ward branches migrate much slower. They reach the sunspot formation latitudes by the start of the following sunspot cycle and continue through this cycle, forming a 22-year `extended solar cycle'. 
\begin{figure}
	% \vspace*{-2.0 cm}
	\begin{center}
		\includegraphics[width=0.6\linewidth]{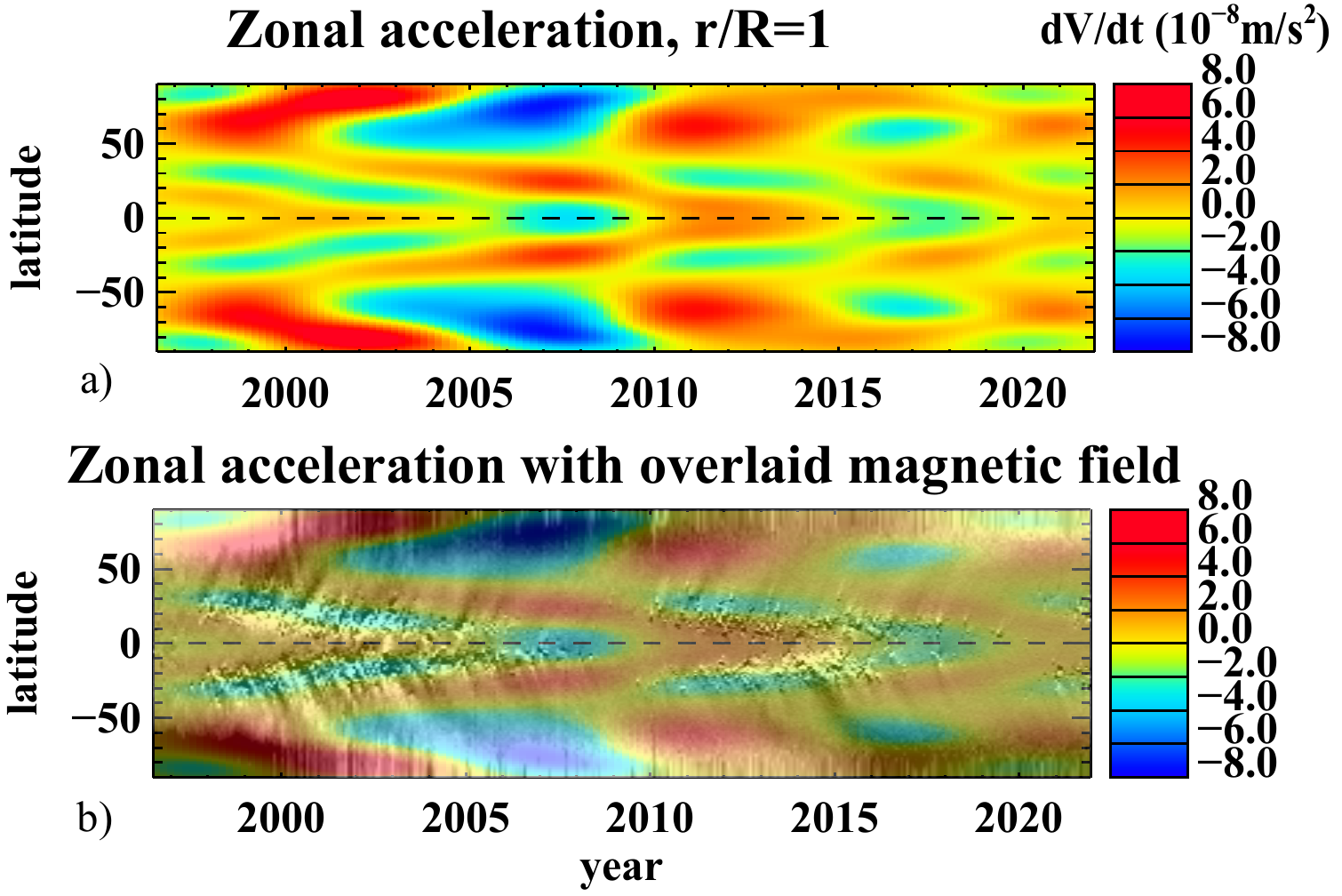} 
		% \vspace*{-1.0 cm}
		\caption{a) Time-latitude diagram of the zonal flow acceleration close to the solar surface from the SoHO/MDI and SDO/HMI data from 1996 to 2022. b) Overlay of the zonal acceleration and the magnetic butterfly diagram for the same period.}
		\label{fig4}
	\end{center}
\end{figure}

The tracking of the torsional oscillation pattern with depth (Fig.~\ref{fig3}) shows that the torsional oscillations originate in the deep convection zone, and thus reflect the internal dynamics associated with the dynamo processes. Zonal acceleration, calculated as the time derivative of zonal velocity (after applying a Gaussian smoothing), (Fig.~\ref{fig4}a), and overlaid the magnetic butterfly diagram (Fig.~\ref{fig4}b), shows that the zones of magnetic flux in the sunspot zone and the polar regions coincide with the zonal deceleration (blue areas). 
     
\begin{figure}[h]
	% \vspace*{-2.0 cm}
	\begin{center}
		\includegraphics[width=0.6\linewidth]{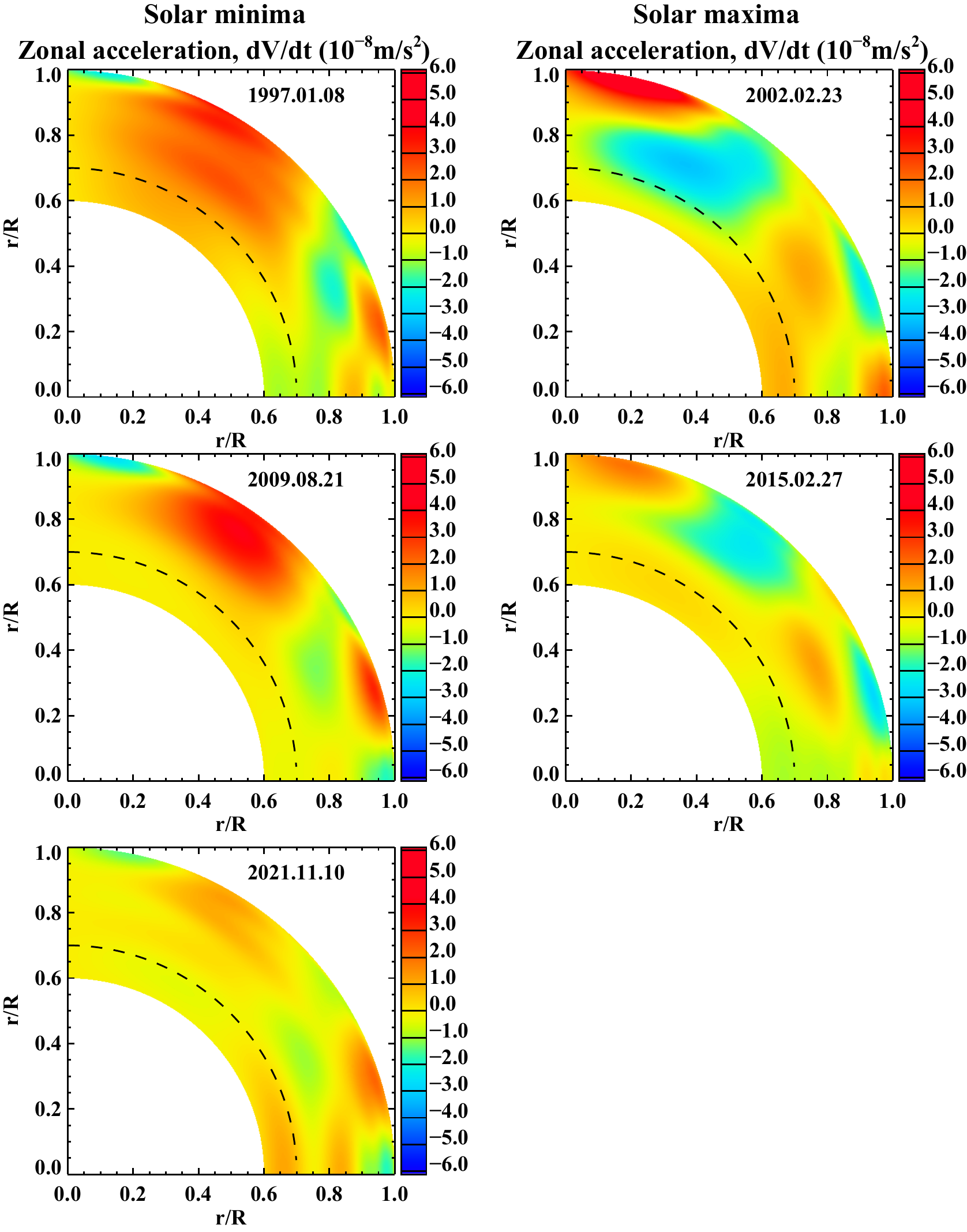} 
		% \vspace*{-1.0 cm}
		\caption{Distributions of the zonal acceleration in the solar interior during the periods corresponding to the solar minima in 1997, 2009, and 2021 (left column) and during the solar maxima in 2002 and 2015 (right columns).}
		\label{fig5}
	\end{center}
\end{figure}

This result indicates that tracking the zonal deceleration with depth may reveal the evolution of internal magnetic fields. The snapshots of the distribution of the zonal acceleration with the depth and latitude during three solar minima and two solar maxima, obtained from the SoHO and SDO data are shown in Figure~\ref{fig5}). During the solar minima, the zonal deceleration regions are extended from the surface mid-latitudes through the convection zone, and also show near-surface deceleration close to the equator. Evidently, these two regions correspond to the start of a sunspot cycle at mid-latitudes and the end of the previous cycle near the equator. During the solar maxima,  the blue deceleration zones in subsurface regions at mid and low latitudes coincide with the sunspot formation zones. In addition, large deceleration regions appear at high latitudes. These regions coincide with the areas of near-polar magnetic field (see Fig.~\ref{fig4}b), and their extension to the bottom of the convection zones indicates the polar magnetic field depends on dynamo processes in a high-latitude zone at the bottom of the convection (the tachocline). 

\begin{figure}[t]
	% \vspace*{-2.0 cm}
	\begin{center}
		\includegraphics[width=0.5\linewidth]{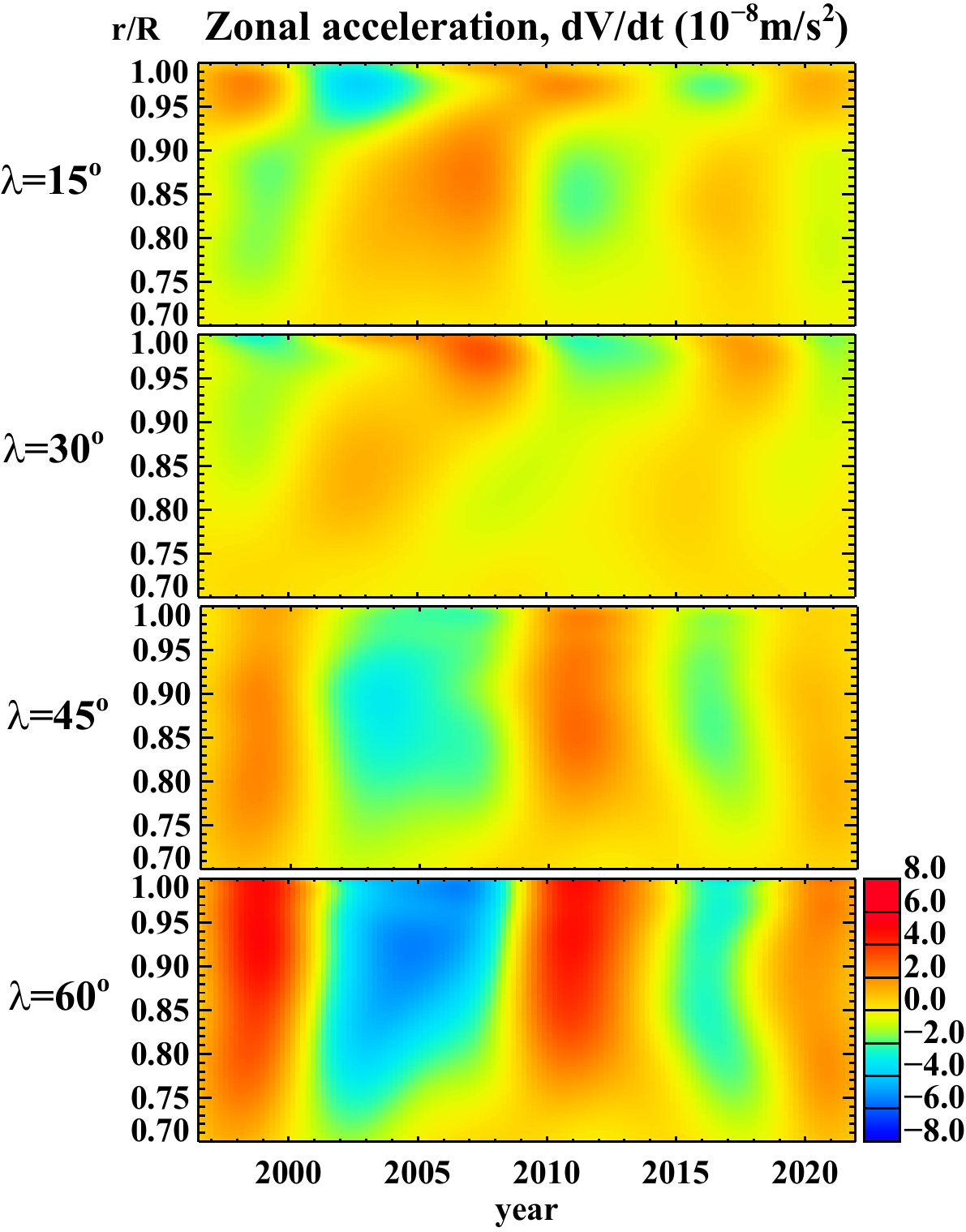} 
		% \vspace*{-1.0 cm}
		\caption{Evolution of the zonal acceleration with depth at four latitudes: 15, 30, 45 and 60 degrees in 1996-2022.}
		\label{fig6}
	\end{center}
\end{figure}

The evolution of the zonal deceleration with depth  (Fig.~\ref{fig6}) shows that both the equator-ward and polar-ward branches  originate at the bottom of the convection zone at about 60 degrees latitude. The polar branch migrates to the surface in 1-2 years, but it takes about 11 years for the equatorial branch to appear at mid-latitudes in the sunspot formation zone. During the following 11 years, it slowly migrates towards the equator. It is intriguing that the migration slows down at the bottom boundary of the subsurface shear layer, at $r/R_\odot \approx 0.95$.

The observed evolution of the subsurface dynamics corresponds to the dynamo wave theory introduced by \citet{Parker1955}, which leads to the following interpretation of the helioseismic observations.
In this theory, the magnetic field evolution is described as `dynamo waves', which represent turbulent diffusion of the magnetic field, amplified by helical motions and differential rotation. The cyclic evolution of solar magnetic fields begins at the bottom of the convection zone at about 60 degrees latitudes. In the high-latitude regions, the Coriolis force acting on flows in seed toroidal magnetic flux tubes twists and amplifies the magnetic field, creating a large-scale poloidal magnetic field (by so-called 'the alpha-effect'). The generated poloidal field quickly migrates to the surface in the high-latitude regions, but on its way to the surface at lower latitudes it is stretched and amplified by the differential rotation (so-called 'the Lambda-effect'). It takes about 11-years for the generated toroidal field to reach the surface at about 30 degrees latitude where it emerges in the form of bipolar sunspot groups, initiating the sunspot butterfly diagram.

 The helioseismology data (Fig.~\ref{fig6}) show that the subsurface shear plays a significant role in the magnetic flux emergence and evolution. The radial gradient of the differential rotation in the near-surface shear layer determines the equator-ward direction of the toroidal field migration in the butterfly diagram \citep{Brandenburg2005,Pipin2011}. The data also indicate a downward migrating branch of the zonal deceleration in the near-surface shear layer, which may be related to the submergence of the magnetic field. Observations show that the surface magnetic field rotates with the speed corresponding to the bottom part of the near-surface shear layer. It means that the magnetic field of sunspots is anchored in this layer. The magnetic time-latitude diagram shows that the decaying toroidal magnetic field is transported by meridional flows and turbulent diffusion to high-latitude regions, where it submerges to the bottom of the convection zone, providing a seed field for the next solar cycle.

In the dynamo-wave theory, the duration of solar cycles is primarily determined by three parameters: the strength of cyclonic motions at the bottom of the convection zone, the differential rotation and turbulent diffusion. Thus, the stability of the 11-year cycles \citep['the solar clock', ][]{Russell2020} can be explained by the long-term stability of the internal dynamics in the deep convection zone. However, the strength of the sunspot cycles changes significantly. The helioseismic measurements show that the amplitude of the zonal acceleration in the convection zone substantially decreased in the past cycle, which is in line with the low sunspot maximum \citep{Kosovichev2019}. 
%The amplitude variations might be related to variations of the magnetic field emergence and dissipation.

In general, the dynamo-wave scenario is supported by helioseismic measurements. However, the underlying non-linear turbulent MHD processes can only be understood through computational modeling.

%\newpage
\section{3D MHD Global-Sun Simulations}

The simulations are performed in a 3D anelastic MHD approximation and in a spherical shell, $0\le \phi \le 2\pi$, $0\le \theta \le \pi$,   $0.61 \geq r/\Rs \leq 0.96$. It includes a transition zone between the radiative and convection zones at $0.7\Rs$ (the tachocline), but does not include the shallow subsurface layers where the anelastic approximation is invalid.
The mathematical model is described in terms of  conservation of mass, momentum, energy, and magnetic flux. The energy equation is written for perturbations of potential temperature, $\Theta$, related to the specific entropy: $s=c_p \ln\Theta+{\rm const}$)  \citep{Ghizaru2010}:
\begin{equation}                                                                                               
	{\bm \nabla}\cdot(\rho_s\bm u)=0, \label{equ:cont}
\end{equation}
\begin{equation}
	\frac{D \bm u}{Dt}+ 2{\bm \Omega} \times {\bm u} =  
	-{\bm \nabla}\left(\frac{p'}{\rho_s}\right) + {\bm g}\frac{\Theta'}
	{\Theta_s} + \frac{1}{\mu_0 \rho_s}({\bm B} \cdot \nabla) {\bm B} \;, \label{equ:mom} 
\end{equation}
\begin{equation}
	\frac{D \Theta'}{Dt} = -{\bm u}\cdot {\bm \nabla}\Theta_e -\frac{\Theta'}{\tau}\;, \label{equ:en} 
\end{equation}
\begin{equation}
	\frac{D {\bm B}}{Dt} = ({\bm B}\cdot \nabla) {\bm u} - {\bm B}(\nabla \cdot {\bm u})  \;,
	\label{equ:in} 
\end{equation}
\noindent
where $D/Dt = \pd/\pd t + \bm{u} \cdot {\bm \nabla}$ is the total (material) derivative, ${\bm u}$ is velocity in the frame rotating with the angular velocity ${\bm \Omega}=\Omega_0(\cos\theta,-\sin\theta,0)$, $p'$ is a pressure perturbation that accounts for both the gas and magnetic pressure, ${\bm B}$ is the magnetic field, and $\Theta'$ is the perturbation of potential temperature  with respect to an ambient state $\Theta_e$ \citep[see Sec.3 of][for a comprehensive discussion]{Guerrero2013}. Furthermore, $\rho_s$  and $\Theta_s$ are the density and potential temperature of the reference state;   ${\bm g}=GM/r^2 \bm{\hat{e}}_r$ is the gravity acceleration, and $\mu_0$ is the magnetic permeability.

The term $\Theta'/\tau$ represents dissipation of the heat flux. %\citep[see section 1.2 and Annexe B in][for details]{Cossette2014}. In this work the parameters of the ambient state are slightly different from those in equation (7) of \cite{Guerrero2013} \S3.1.
For the models including a tachocline, we use the polytropic model with indexes $m_r=2$ in the radiative zone and $m_{\rm cz}=1.499978$ in the convection zone (this value is just below the convective instability threshold). The transition between the radiative and convection zone is modeled by a smooth polytropic index variation with width $w_t=0.015\Rs$. The relaxation time of the potential temperature perturbations is chosen $\tau=1.036\times10^8$ s ($\sim 3.3$ yr).

The equations are solved numerically using the EULAG-MHD code  
\citep{Smolarkiewicz2013,Guerrero2013}. 
The time evolution is calculated using a special semi-implicit method based on a 
high-resolution, non-oscillatory forward-in-time advection scheme MPDATA 
\citep[Multidimensional Positive Definite Advection Transport Algorithm;][]{Smolarkiewicz2006}. 
The truncation terms in MPDATA  evince viscosity comparable to the explicit sub-grid scale viscosity used in Large-Eddy Simulation (LES) models. Thus, the results of MPDATA are often interpreted as an implicit version of the LES method to account for unresolved turbulent transport. The boundary conditions are impermeable stress-free for the velocity field. The convective heat flux is set to zero at the top, and the flux divergence is zero at the bottom. In addition, it is assumed that the magnetic field is radial at the boundaries.

\begin{figure}[t]
	% \vspace*{-2.0 cm}
	\begin{center}
		\includegraphics[width=0.75\linewidth]{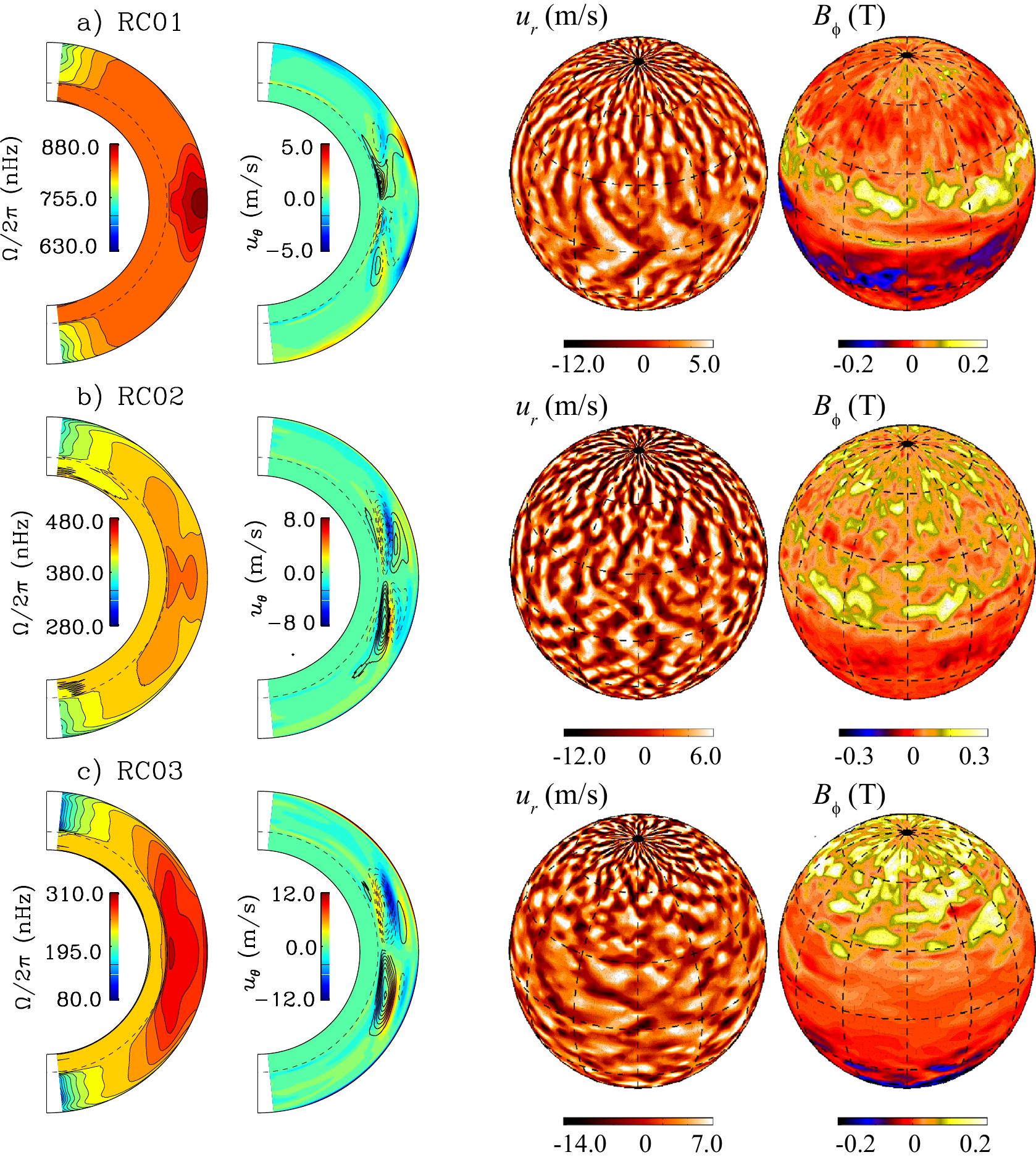} 
		% \vspace*{-1.0 cm}
		\caption{From left to right: mean profiles of the rotation rate, $\Omega/2\pi$, and meridional velocity, $v_\theta$ 
			and snapshots at $r=0.95\Rs$ of the vertical velocity, $u_r$, and the toroidal field, 
			$B_{\phi}$, for the simulated models a) RC01: $\Omega=2\Omega_\odot$, b) RC02: $\Omega=\Omega_\odot$, and c) RC03: $\Omega=0.5\Omega_\odot$. In the meridional circulation panels,
			continuous (dashed) lines correspond to clockwise (counterclockwise) circulation.
		%	The background filled contours show the latitudinal velocity, $\mean{u}_{\theta}$.
			The differential rotation and meridional circulation profiles are calculated 
			from the mean azimuthal values averaged over $\sim 3$ years.}
		\label{fig7}
	\end{center}
\end{figure}

Figure \ref{fig7} shows 
the differential rotation (left column) and meridional circulation (second column) profiles of the computational models with different rotation rate: a) RC01 - $\Omega=2\Omega_\odot$, b) RC02 - $\Omega=\Omega_\odot$, c) RC03 - $\Omega=0.5\Omega_\odot$. All these models exhibit the solar-type differential rotation -- low-latitude regions rotate faster than the high-latitude regions. Model ~RC01 has the lowest Rossby number; thus, it has the strongest influence of the Coriolis force. In this model, the latitudinal differential rotation is concentrated in low-latitude regions. In the other two models, it is distributed over latitudes. 

%This force tends to homogenize the rotation of the convection zone with that of the radiative zone. This happens in a region between $\sim 5^{\circ}$ and $\sim 80^{\circ}$ latitude. However, zones with some radial and latitudinal shears are developed at the equatorial and polar latitudes. In model ~RC02, both radial and latitudinal differential rotations develop in larger zones. The intermediate latitudes are in iso-rotation with the radiative zone.

%The transport of angular momentum to the radiative zone makes the stable layer to rotate, on average (cf. Fig. \ref{fig.df2}b), faster than the frame. Hence, the equatorial acceleration of the convection zone in the model RC02 is not as pronounced as in the model CZ02. In the former case the angular velocity is around 440nHz, while in the latter case it reaches 500nHz. Consequently, the latitude of iso-rotation with the frame is moved up to $\sim 70^o$.  In the model  CZ02 it is located at $\sim 50^o$ (the yellow filled contours in Figs. \ref{fig.df1}b and \ref{fig.df2}b correspond approximately to the frame rotation rate, $\Omega_0$).

The magnetic field is generated by the turbulent dynamo from random initial perturbations. It affects the mean differential rotation (making it solar-like). In addition, the magnetic feedback on the flow generates zonal flows (torsional oscillations). The latitudinal flow structure resembles the observations, but the zonal flow velocity amplitude is several times larger than the observed velocity. The torsional oscillations are driven by the magnetic torque and by a modulation of the latitudinal angular momentum transport mediated by the meridional circulation  \citep{Guerrero2016}. 

The model reproduces the near-surface rotational shear layer (NSSL). However, it is mainly pronounced at higher latitudes, which contradicts the helioseismic observations showing that the NSSL is almost uniform in latitude.
The meridional circulation is multicellular in all models. However, as expected, the meridional velocity in model RC03 (with the lower rotation rate) is higher than in the other models. Also, it exhibits a dominant counterclockwise cell in the Northern hemisphere and a clockwise cell in the Southern hemisphere. Noteworthy, a low amplitude poleward flow develops in the upper part of the convection zone at latitudes $>30^{\circ}$ in all these models (see in second column of  of Fig. \ref{fig7}.

The two right columns of Fig.~\ref{fig7} show snapshots of the radial velocity and magnetic field at the top boundary, $r=0.96\Rs$. Evidently, the scale of convection varies with latitude: smaller convective structures are formed at higher latitudes. At low latitudes, the structures are elongated, resembling `banana’ cells. Such convection structuring is not observed on the surface. Therefore, in future work, it is important to investigate the spectrum of subsurface solar convection by helioseismology to check the model predictions. 

The simulations show that the dynamo processes strongly depend on the rotation rate. In the fast-rotating model, RC01, the dynamo-generated magnetic field oscillates in amplitude but does not show clear polarity reversals. The field's topology consists of wreaths of the toroidal field of opposite polarity in the Northern and Southern hemispheres. Magnetic cycles with a full period of  $\sim 30$~yr are developed in model RC02. However, in this model, the toroidal magnetic field polarity is the same in both hemispheres, contrary to the observations. The amplitude of the surface magnetic field correlates directly with the rotation rate. The mean field strength reaches $10^4$~G in the tachocline because of the strong radial shear.

%\newpage
\section{Mean-Field Models of the Solar Dynamics and Dynamo}

The mean-field model \citep{Pipin2018}  describes the evolution of the mean axisymmetric velocity, magnetic field, and entropy. The mean axisymmetric velocity is represented in terms of the angular velocity, $\Omega$, and the meridional circulation velocity $\mathbf{\overline{U}}^{m}$: $\mathrm{\mathbf{\overline{U}}=\mathbf{\overline{U}}^{m}+r\sin\theta\Omega\hat{\mathbf{\boldsymbol{\phi}}}}$,
where $\boldsymbol{\hat{\phi}}$ is the azimuthal unit vector.
Similar to the 3D simulations, we employ the anelastic approximation. The full MHD system in the anelastic approximation includes the conservation of the angular momentum, the equation for the azimuthal component of vorticity ${\overline{\omega}=\left(\boldsymbol{\nabla}\times\overline{\mathbf{U}}^{m}\right)_{\phi}}$, the heat transport equation in terms of mean entropy $\overline{\mathrm{s}}$, and the induction equation for the large-scale azimuthal magnetic field $\overline{\mathbf{B}}$:
\begin{eqnarray}
	\frac{\partial}{\partial t}\overline{\rho}r^{2}\sin^{2}\theta\Omega & = & -\boldsymbol{\nabla\cdot}\left(r\sin\theta\overline{\rho}\left(\hat{\mathbf{T}}_{\phi}+r\sin\theta\Omega\mathbf{\overline{U}^{m}}\right)\right)
	%	\\ & + & 
	+ \boldsymbol{\nabla\cdot}\left(r\sin\theta\frac{\overline{\mathbf{B}}\overline{B}_{\phi}}{4\pi}\right) \label{eq:angm}
\end{eqnarray}
\begin{eqnarray}
	{\frac{\partial\omega}{\partial t}} =  \mathrm{r\sin\theta\boldsymbol{\nabla}\cdot\left(\frac{\hat{\boldsymbol{\phi}}\times\boldsymbol{\nabla\cdot}\overline{\rho}\hat{\mathbf{T}}}{r\overline{\rho}\sin\theta}-\frac{\mathbf{\overline{U}}^{m}\overline{\omega}}{r\sin\theta}\right)}
	+  \mathrm{r}\sin\theta\frac{\partial\Omega^{2}}{\partial z}-\mathrm{\frac{g}{c_{p}r}\frac{\partial\overline{s}}{\partial\theta}}+F_{L}^{(p)}\label{eq:vort} 
\end{eqnarray}
\begin{equation}
	\overline{\rho}\overline{T}\left(\frac{\partial\overline{\mathrm{s}}}{\partial t}+\left(\overline{\mathbf{U}}\cdot\boldsymbol{\nabla}\right)\overline{\mathrm{s}}\right)=-\boldsymbol{\nabla}\cdot\left(\mathbf{F}^{c}+\mathbf{F}^{r}\right)-\hat{T}_{ij}\frac{\partial\overline{U}_{i}}{\partial r_{j}}-\boldsymbol{\boldsymbol{\mathcal{E}}}\cdot\left(\nabla\times\overline{\boldsymbol{B}}\right),\label{eq:heat}
\end{equation}
\begin{equation}
	\frac{\partial\overline{\mathbf{B}}}{\partial t}
	=\boldsymbol{\nabla}\times\left(\boldsymbol{\mathcal{E}}+\mathbf{\overline{U}}\times\overline{\mathbf{B}}\right),\label{eq:mfe-1}
\end{equation}
where $\mathrm{\partial/\partial z=\cos\theta\partial/\partial r-\sin\theta/r\cdot\partial/\partial\theta}$ is the gradient along the axis of rotation, $\hat{\mathbf{T}}$ is the turbulent stress tensor that includes
small-scale fluctuations of velocity and magnetic field: 
\begin{equation}
	\hat{T}_{\mathrm{ij}}=\overline{u_{i}u_{j}}-\frac{1}{4\pi\overline{\rho}}\left(\overline{b_{i}b_{j}}-\frac{1}{2}\delta_{ij}\overline{\mathbf{b}^{2}}\right),\label{eq:stres}
\end{equation}
$\mathbf{u}$ and $\mathbf{b}$ are the turbulent fluctuating velocity
and magnetic field, $\overline{\rho}$ is the mean density, $\overline{T}$ - the mean
temperature, $\mathbf{F}^{c}$ is the eddy convective flux, $\mathbf{F}^{r}$
is the radiative flux, $F_{L}^{(p)}$ is the poloidal component of the large-scale Lorentz force, $\boldsymbol{\mathcal{E}}$
is the mean electromotive force in the form: 
\begin{equation}
	\mathcal{E}_{i}=\left(\alpha_{ij}+\gamma_{ij}\right)\overline{B}_{j}-\eta_{ijk}\nabla_{j}\overline{B}_{k}.\label{eq:EMF-1-1}
\end{equation}
where symmetric tensor $\alpha_{ij}$ models generation of the large-scale
magnetic field by the $\alpha$-effect, which includes kinetic and magnetic helicities; anti-symmetric tensor $\gamma_{ij}$
controls the mean drift of the large-scale magnetic fields in turbulent
medium; the tensor $\eta_{ijk}$ describes the anisotropic turbulent
diffusion \citep[for more details, see ][]{Pipin2018}.

\begin{figure}[t]
	% \vspace*{-2.0 cm}
	\begin{center}
		\includegraphics[width=0.7\linewidth]{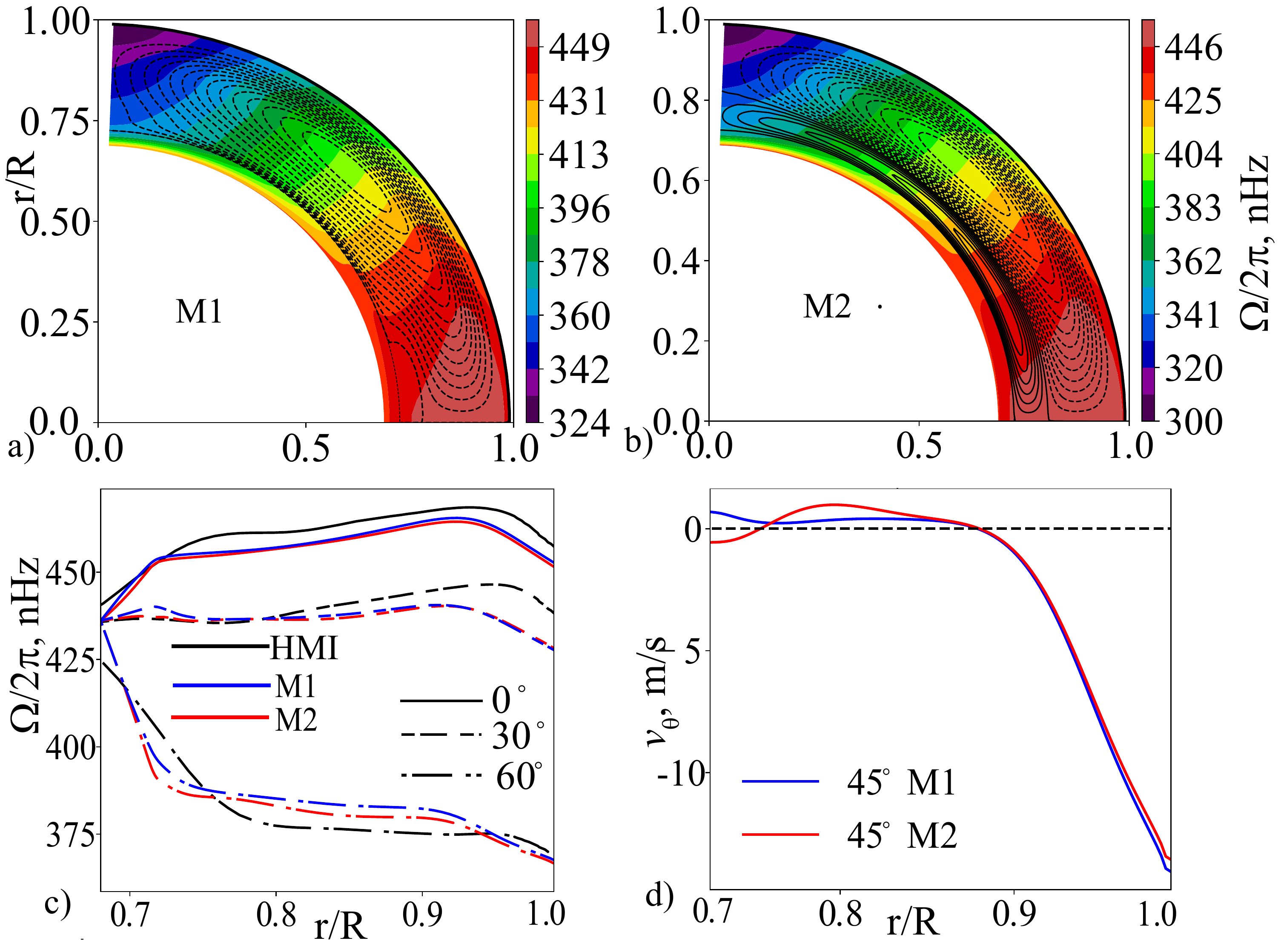} 
		% \vspace*{-1.0 cm}
		\caption{The angular velocity and meridional circulation distributions
			for model M1, which has a single-cell meridional circulation, b) model M2 with a double-cell meridional circulation c)
			radial profiles of the angular velocity at 0, 30 and 60 degrees latitudes
			{( blue for model M1, red - M2), and obtained from the SDO/HMI
				helioseismology data archive (black);} d) radial profiles of the meridional
			circulation velocity at latitude 45$^{\circ}$.}
		\label{fig8}
	\end{center}
\end{figure}

Traditional mean-field dynamo models prescribe the differential rotation and meridional circulation and solve only the induction equation (Eq.~\ref{eq:mfe-1}). But, this model solves the full system of the MHD equations, and reproduces the differential rotation, meridional circulation, magnetic field and their variations with the solar cycles. Figure \ref{fig8} shows the differential rotation and meridional circulation for two models M1 and M2. Their differential rotation profiles are in good agreement with the helioseismic measurements from the SDO/HMI instrument. However, the subsurface shear layer in the models is somewhat broader and less steep compared to the observations (Fig.~\ref{fig8}c).  Models M1 and M2 are calculated for identical conditions except that Model M2 takes into account the radial inhomogeneity of the Coriolis number, which depends on the convective turnover time. This effect results in a weak secondary meridional circulation cell at the bottom of the convection zone \citep[Fig.~\ref{fig8}b,d; ][]{Pipin2018}.

\begin{figure}[t]
	% \vspace*{-2.0 cm}
	\begin{center}
		\includegraphics[width=0.7\linewidth]{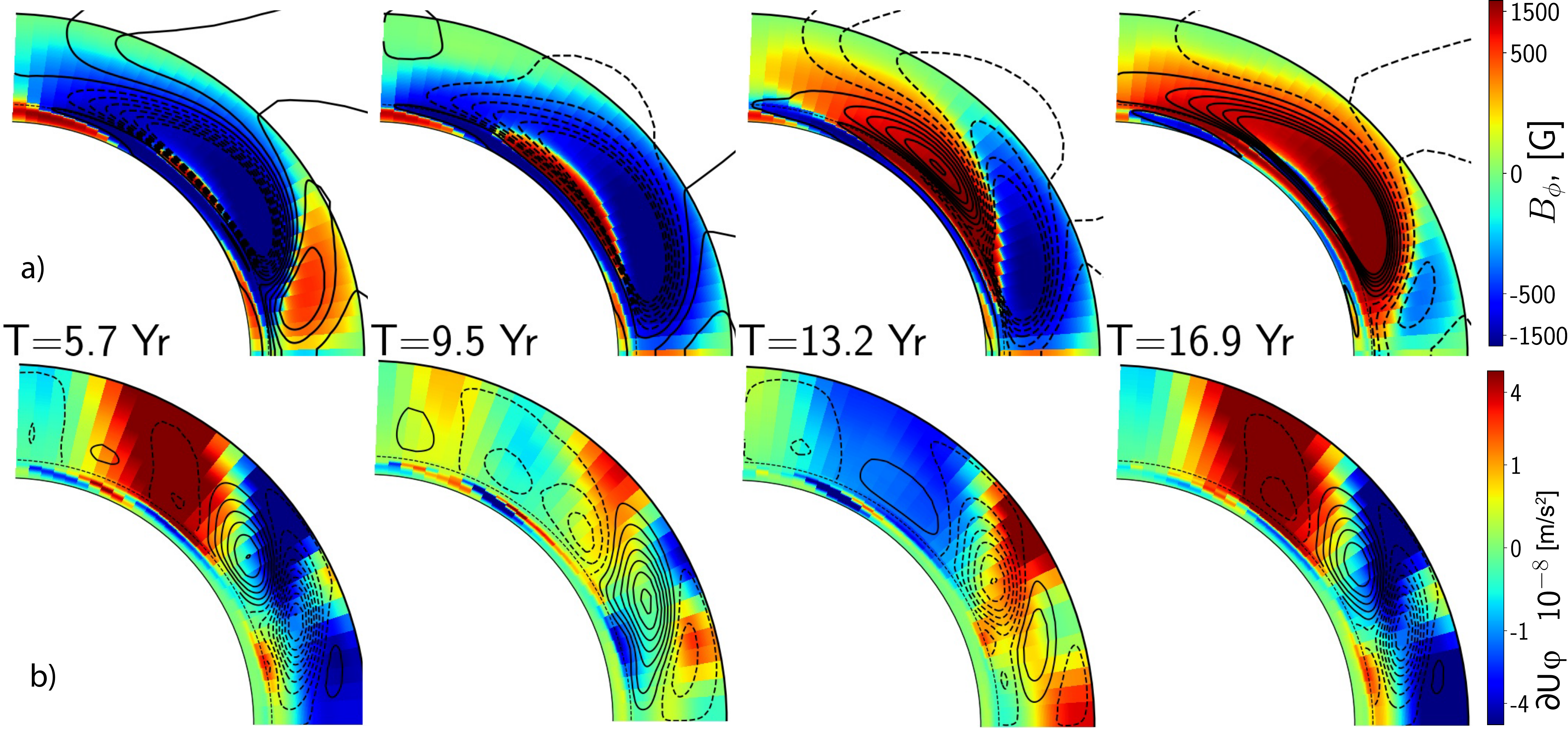} 
		% \vspace*{-1.0 cm}
		\caption{Model M1. Snapshots for a half of dynamo cycle of: a)
			the toroidal magnetic field (background image) and streamlines of
			the poloidal field (contours); b) variations of the zonal acceleration
			(background image) and the azimuthal force caused by variations of
			the meridional circulation (contour lines are plotted in
			the range $\pm50$~m/s$^{2}$).}
		\label{fig9}
	\end{center}
\end{figure}

\begin{figure}[t]
	% \vspace*{-2.0 cm}
	\begin{center}
		\includegraphics[width=0.9\linewidth]{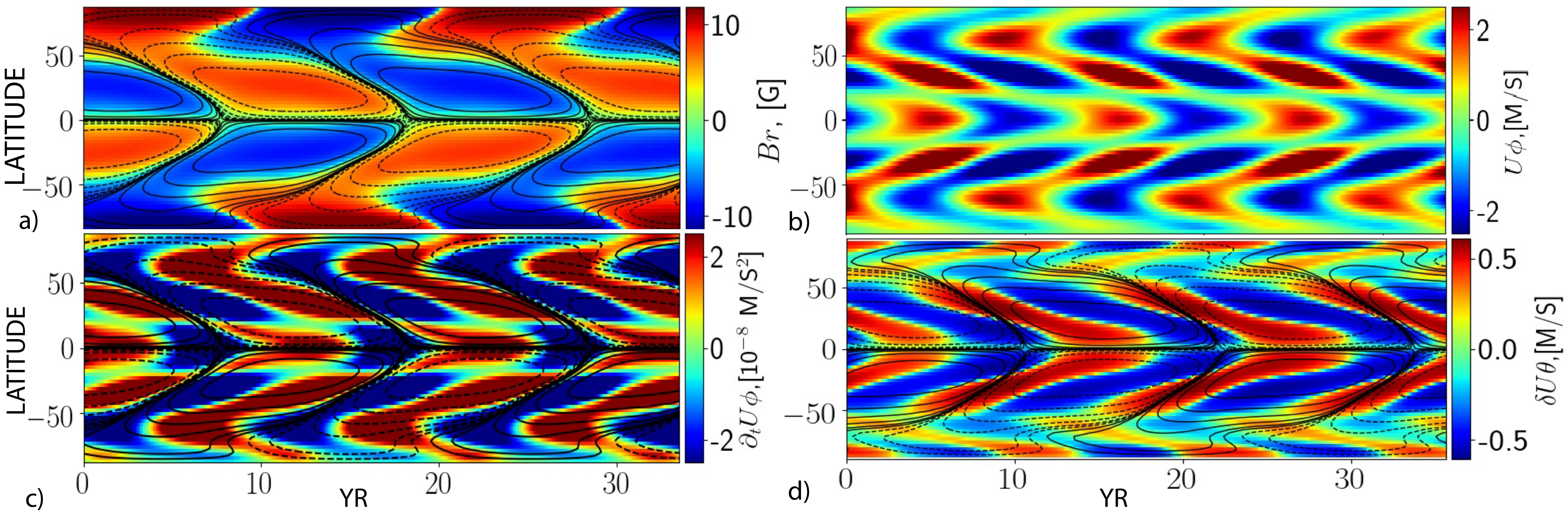} 
		% \vspace*{-1.0 cm}
		\caption{Model M1. a) Time-latitude diagram of the radial magnetic
			field at the surface (background image) and the toroidal magnetic
			field at $r_{s}=0.9\Rs$,
			()contour lines are plotted in range of $\pm$1kG with an exponential
				decrease to the low values of $\pm$4G; b) time-latitude
			diagram of the torsional oscillations (background image) at the surface;
			c)  time-latitude diagram of the zonal
			flow acceleration (background image) at the surface and the toroidal
			magnetic field in the subsurface shear layer (same as panel (a)); d) variations of the meridional circulations at the surface (positive values correspond to
			the poleward flow).}
		\label{fig10}
	\end{center}
\end{figure}

Figure \ref{fig9} shows snapshots of the large-scale magnetic
field and zonal acceleration in the convection zone for half of the solar cycle for model M1.
A new dynamo cycle starts near the bottom of the convection zone {
	at about 60 degrees latitude} in the region subjected to  zonal
{ deceleration}. The acceleration pattern in the upper part of the
convection zone drifts toward the equator following the 
the dynamo wave propagation (see, Fig.~\ref{fig9}a,b). 
We see that in the latitude range
from 50$^{\circ}$ to 60$^{\circ}$, where the extended dynamo mode
is initiated, the acceleration is provided by the inertia force.
The polar branch of the torsional oscillations in model M1 is
due to the effects of the Lorentz force and variations of the meridional
circulation. 

%Figures~\ref{czM1}d,e shows that the zonal deceleration is ahead of the toroidal magnetic field. 

%due to the combined effect of the large-scale Lorentz force, $F_{L}^{(t)}$, and the dynamo induced $\Lambda$-effect, $F_{H}$. Notably, the $F_{H}$ force migrates closer to the equator than the $F_{L}^{(t)}$ force. The forces $F_{U}$ (Fig.~\ref{czM1}b) and $F_{I}$ (Fig.~\ref{czM1}e) are opposite in sign, and they have similar latitudinal structures.

%, $F_{I}$ (Figs.~\ref{czM1}b and e). Also, we see that, during a half of the extended torsional oscillation cycle, effects of $F_{U}$, $F_{L}$ and $F_{H}$ are synchronized in the subsurface layer of the convection zone. 

Figure \ref{fig10} shows the radial magnetic field evolution in model M1 at the surface and the corresponding evolution of dynamo-induced zonal variations of the rotational velocity and zonal acceleration. The large-scale toroidal magnetic field, shown at the bottom of the near-surface shear layer at $r\sim 0.9R_\odot$, varies with the magnitude of 1.5~kG. In the subsurface shear layer, the dynamo wave of the toroidal magnetic field starts at about $60^{\circ}$ latitude, approximately 1-2 years after the end of a previous activity cycle. The toroidal magnetic field strength at this latitude is about 4~G. The wave propagates toward the equator in $\sim22$   years. The polar and equatorial branches of the dynamo waves almost completely overlap in time. The evolution of the radial magnetic field also reveals the extended cycle and agrees with the observational results \citep{Stenflo2012}.

The dynamo wave forces variations of the angular velocity and meridional
circulation. It is seen in Fig.~\ref{fig10}b that the induced zonal
acceleration is $\sim2-4\times10^{-8}~$m$\,$s$^{-2}$, which is
in agreement with the observational results of \citet{Kosovichev2019}.
However, the individual force contributions are stronger than their combined action by more than an order of magnitude. Another interesting
finding is that two components of the azimuthal force show the extended
22-year modes. These forces are associated with variations of the
meridional circulation, and the inertial force. The polar branch of the torsional oscillations in the models is due to the effects of the Lorentz force and variations of the meridional circulation. 

The model shows weak meridional circulation variations in the main part of the convection zone, where its magnitude is about 10–20 cm/s. The surface variations of the meridional circulation in the dynamo cycle are about 1 m/s (Fig.~\ref{fig10}d). They correspond to meridional flows converging towards the activity belts, in agreement with results of local helioseismology \citep{Komm2012,Zhao2014,Kosovichev2016}.

The results show that the large-scale toroidal magnetic field results in a reduction of the convective flux. This phenomenon, called the magnetic shadow effect, was discussed earlier by, e.g., \citet{Brandenburg1992} and \citet{Pipin2004}, and is usually considered in the problem of the solar-cycle luminosity variation. In the model, the magnetic shadow effect induces variations of the latitudinal gradient of the mean entropy. This results in perturbation of the Taylor-Proudman balance and variations of the meridional circulation. In agreement with other studies \citep[e.g.,][]{Durney1999,Rempel2006,Miesch2011}, the variations are concentrated near the boundaries of the convection zone.

Thus, the model shows that the torsional oscillations are driven by a combination of magnetic field effects acting on turbulent angular momentum transport and the large-scale Lorentz force. We find that the 22-year `extended' cycle of the torsional oscillations results from a combined effect of the overlap of subsequent magnetic cycles and magnetic quenching of the convective heat transport. This quenching results in variations of the meridional circulation. The variations of the meridional circulation together with other drivers of the torsional oscillations maintain their migration to the equator forming the 22-year extended cycle.

\section{Using Computational Models for Prediction of Solar Dynamics and Activity}
\begin{figure}[t]
	% \vspace*{-2.0 cm}
	\begin{center}
		\includegraphics[width=0.65\linewidth]{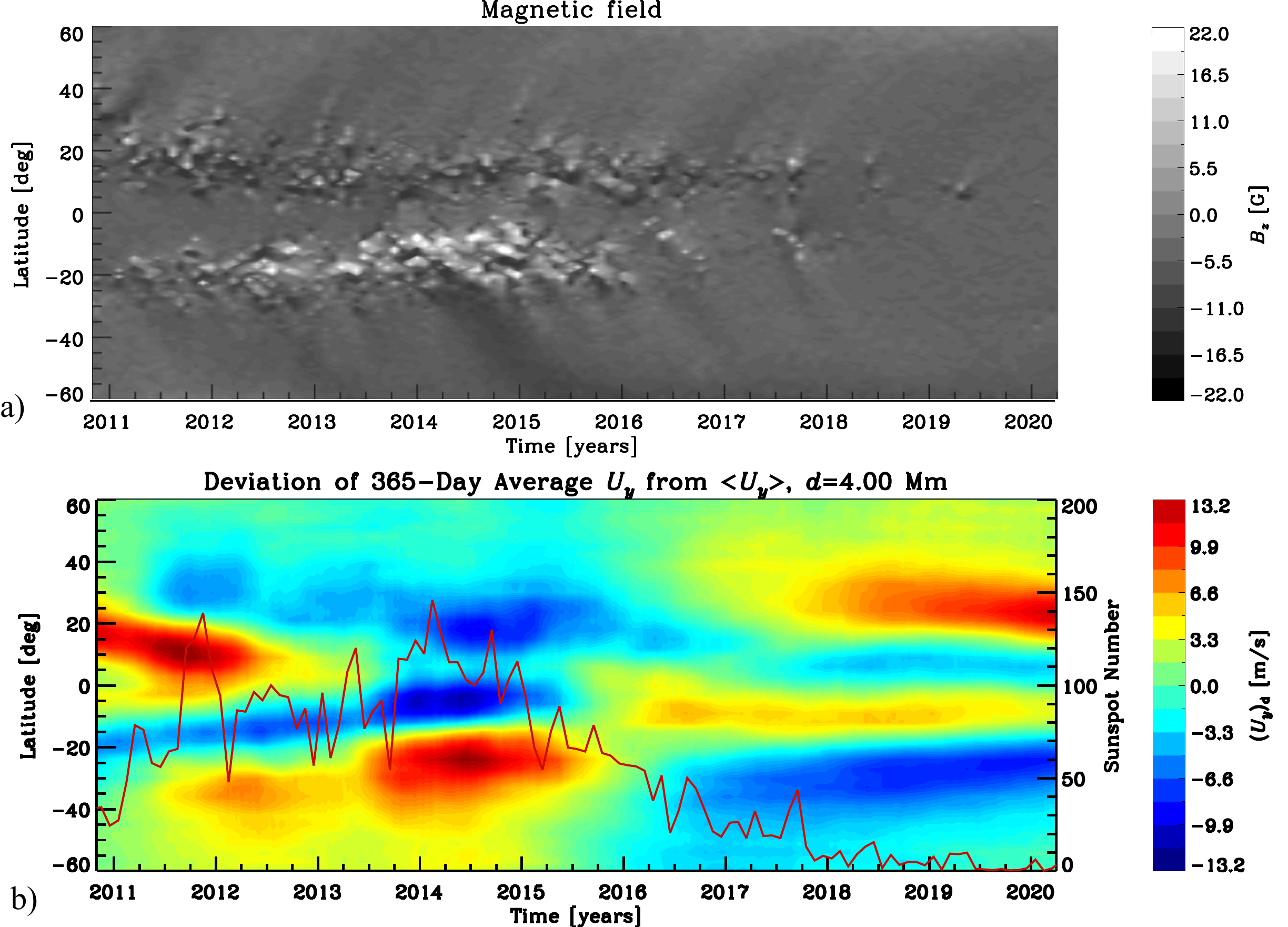} 
		% \vspace*{-1.0 cm}
		\caption{Time-latitude diagrams: a) monthly averaged Bz, b) time-latitude diagrams representing the deviation of the 365-day running average of the meridional velocity, Uy, from its mean over the whole interval at depth 4 Mm}
		\label{fig11}
	\end{center}
\end{figure}

Observations established that the strength of the polar magnetic field during solar minima correlates with the next sunspot maximum \citep{Schatten1978}. The mean-field dynamo model explains this relationship. It also predicts a correlation between the strength of the zonal acceleration at the base of the convection zone at high latitudes (where according to the model and interpretation of the helioseismic measurements, the solar cycle originates) and the amplitude of the following solar cycle \citep{Pipin2020}. The predicted correlation depends on the mechanism that causes the variations of the solar cycles, which is not yet established. Therefore, two types of dynamo models were considered: models with regular variations of the  $\alpha$-effect, and models with stochastic fluctuations, simulating `long-memory' and `short-memory' variations of magnetic activity. It is found that torsional oscillation properties, such as the zonal acceleration, correlate with the magnitude of the subsequent cycles with a time lag of 11-20 yr. The correlation sign and the time-lag depend on the depth and latitude where the torsional oscillations are measured and on the properties of the long-term variations of the dynamo cycles. The strongest correlations with the future cycles are found for the zonal acceleration at the base of the convection zone at high ($\sim 60$ deg) latitudes. The modeling results demonstrate that helioseismic observations of the torsional oscillations can be useful for advanced prediction of the solar cycles, 1-2 sunspot cycles ahead. However, uncertainties remain because the mechanism of the solar cycles variations is not known, and the results for short- and long-memory models differ. The continuing helioseismic observations will allow us to test these predictions.

The extended solar cycle of the migrating meridional flow pattern is an important model prediction. It can be tested using already available local helioseismology data. Previous data analyses found variations of the meridional circulation associated with near-surface flows converging around active regions. However, the model predicts that such variations of the meridional circulation are part of the global dynamo processes. Similarly to the torsional oscillations, they extend through periods of low sunspot activity, forming the extended solar cycle pattern (Fig.~\ref{fig10}d).

This prediction is confirmed by analysis of subsurface flow maps obtained by the Time-Distance \citep{Getling2021} and Ring-Diagram \citep{Komm2021}  techniques. Figure~\ref{fig11}  shows the time-latitude diagram of the surface magnetic field and the subsurface South-North velocity variations, obtained after subtracting the mean velocity at 4 Mm depth from May 2010 to September 2020. The velocity variations change the sign across the equator, forming bands of flows converging towards the zones of sunspot formation during Solar Cycle 24, in 2010-16. The flow migration pattern with latitude is similar to the magnetic butterfly diagram. Still, it continues (with a slight latitudinal shift) in 2016-20 when there was no significant surface magnetic activity. Thus, the variations of the meridional circulation form the extended solar-cycle pattern, which is very similar to the extended cycle of the torsional oscillations. Figure~\ref{fig11}b reveals the flow velocity increases during periods of high solar activity. These enhancements correspond to the previously discovered converging flows around active regions. Thus, the variations of the meridional circulation consist of two components: flows converging around active regions and a 22-year extended solar-cycle component associated with the solar dynamo. Due to the North-South asymmetry of active regions, the first component is asymmetric (causing apparent cross-equatorial meridional flows). The second component is mostly North-South symmetrical. Such behavior suggests that the global dynamo processes are mostly North-South symmetrical, and that the hemispheric asymmetry of active regions on the surface is associated with the process of magnetic flux emergence.

\section{Discussion}

In this paper, we presented some examples of computational heliophysics models that help understand the complex turbulent dynamics inside the Sun and the origin of the Sun’s magnetic activity and cycles. The processes beneath the visible surface of the Sun can be observed only indirectly by helioseismic techniques. However, the helioseismic inversion of noisy measurements of oscillation frequencies and acoustic travel times are intrinsically ill-posed and cannot provide a unique solution. The computational solar acoustics, presented in Sec. 3, provides an important tool for directly testing theoretical models of the solar dynamics by forward modeling the observational data. In particular, it has shown that the current travel-time measurements are consistent with the single- and double-cell structure of the meridional circulation and that the helioseismic data can establish only an upper constraint on the potentially existing deep secondary cell. This result shifts the focus of the current debates on the meridional circulation structure. Further development of the forward helioseismic modeling will help to determine the strength and spectrum of subsurface convective flows, the structure of the tachocline and its variations with the solar cycle, the evolution of the torsional oscillations and their relationship to internal magnetic fields, and solve other critical problems of solar activity.

Measurements of the internal rotation of the Sun during the last two solar cycles revealed zones of fast and slow rotation (`torsional oscillations'), the patterns of which migrating with latitude and depth form `extended' 22-cycles. Our analysis showed that the photospheric magnetic field distribution on the time-latitude diagrams coincides with the regions of zonal deceleration near the surface. This suggests that the tracking of the zonal deceleration in the convection zone may reveal the evolution of internal magnetic fields. Indeed this tracking showed migrating patterns of the zonal deceleration resembling dynamo waves predicted by Parker’s dynamo theory \citep{Parker1955}. Our results also showed that the near-surface rotational shear layer plays a crucial role in the formation of the magnetic butterfly diagram (`shaping the solar cycle'), as suggested by \citep{Brandenburg2005}.

Understanding the observed dynamics can be achieved only through detailed computational models. Unfortunately, the currently available computational resources do not allow us to perform realistic MHD simulations of the whole convection zone from the tachocline to the solar surface. Most global simulations are performed in the anelastic approximation, which neglects the flow compressibility and excludes the surface and near-surface layers. In addition, the resolution of such simulations is not sufficient for describing the whole spectrum of essential turbulent scales. The inability to model the multi-scale turbulent convection is probably the primary reason why these models do not reproduce the observed differential rotation, meridional circulation, and the butterfly diagram. Nevertheless, they provide insight into the complex 3D interaction of turbulent flows and magnetic fields in the highly stratified and rotating convection zone. In particular, we presented the simulation results using a computational model which attempts to describe unresolved turbulent scales using the so-called Implicit Large-Eddy Simulations (ILES) approach. The model reproduces the solar-type differential rotation, tachocline formation, and the near-surface shear layer, albeit only qualitatively.

The quantitative agreement with observations can be obtained in the mean-field MHD approximation, which separates small-scale turbulence from large-scale flows and magnetic fields. In this approximation, the equations for fluctuating turbulent properties are solved, assuming that the large-scale flows and magnetic fields are slowly evolving, and prescribing the turbulence spectrum. The turbulent transport coefficients are then expressed in terms of the large-scale parameters and substituted in the equations describing the large-scale evolution. The mean-field theory has been developed to a high degree of sophistication. In addition to the usual turbulent diffusion transport, it predicts non-diffusive transport of magnetic field and momentum, so-called alpha- and Lambda-effects. The non-diffusive turbulent transport explains the differential rotation and the cyclic evolution of dynamo-generated magnetic fields. With an appropriate choice of free parameters describing the strength and spectrum of the small-scale turbulence, the mean-field models reproduce the observed differential rotation, meridional circulation, and the global magnetic field evolution.

We presented a recently-developed 2D mean-field MHD model, which, in addition, reproduced the evolution of the migrating zonal flows – torsional oscillations and zonal acceleration in close agreement with the observations. It is essential that the model reproduced and explained the 22-year cyclic evolution of the torsional oscillations, which had been a puzzle for many years. The model analysis showed the ‘extended’ cycle is caused by magnetic quenching of the turbulent heat transport, which affects the meridional circulation and the angular-momentum balance. Remarkably, the model predicted the extended cycle of the meridional circulations, confirmed by the helioseismic observations.

Currently, among the computational models, only the mean-field MHD model can reproduce the basic features of the solar dynamics and activity. However, the limitation of this model is that it is two-dimensional and thus cannot describe magnetic field structures emerging on the solar surface in the form of compact bipolar active regions. The formation and evolution of these structures may significantly affect the global evolution of magnetic fields and subsurface flows. Therefore, initial 3D mean-field MHD models describing emerging-flux effects are being developed \citep{Pipin2021}.

Further progress in our understanding of the global dynamics and dynamo of the Sun will be based on a synergy of helioseismic observations with theoretical and computational models. The computational solar acoustics can improve the accuracy of helioseismic inferences of the differential rotation, meridional circulation, and large-scale subsurface flows and their solar-cycle variations. It also provides a tool for the forward modeling of helioseismic observables and direct testing of MHD models. The 3D MHD modeling will be focused on a better understanding of the multi-scale coupling of the turbulent and large-scale flows and fields, using helioseismic constraints and advantages of the mean-field theory and computational models.

%\newpage
Acknowledgments. The work was partially supported by NASA grants NNX14AB70G, 80NSSC20K1320, and 80NSSC20K0602. VVP acknowledges the support of the Ministry of Science and Higher Education of the Russian Federation (Subsidy No.075-GZ/C3569/278).
\bibliographystyle{astron}
%\bibliography{IAUS362_1}

\begin{thebibliography}{}
	
	\bibitem[\protect\astroncite{{Bekki} and {Yokoyama}}{2017}]{Bekki2017}
	{Bekki}, Y. and {Yokoyama}, T. 2017,
	\newblock {\em \apj} {\bf 835(1)}, 9
	
	\bibitem[\protect\astroncite{{B{\"o}ning} et~al.}{2017}]{Boning2017}
	{B{\"o}ning}, V. G.~A., {Roth}, M., {Jackiewicz}, J., and {Kholikov}, S. 2017,
	\newblock {\em \apj} {\bf 845(1)}, 2
	
	\bibitem[\protect\astroncite{{Brandenburg}}{2005}]{Brandenburg2005}
	{Brandenburg}, A. 2005,
	\newblock {\em \apj} {\bf 625(1)}, 539
	
	\bibitem[\protect\astroncite{{Brandenburg} et~al.}{1992}]{Brandenburg1992}
	{Brandenburg}, A., {Moss}, D., and {Tuominen}, I. 1992,
	\newblock {\em \aap} {\bf 265}, 328
	
	\bibitem[\protect\astroncite{{Charbonneau} and
		{Smolarkiewicz}}{2013}]{Charbonneau2013}
	{Charbonneau}, P. and {Smolarkiewicz}, P.~K. 2013,
	\newblock {\em Science} {\bf 340(6128)}, 42
	
	\bibitem[\protect\astroncite{{Chen} and {Zhao}}{2017}]{Chen2017}
	{Chen}, R. and {Zhao}, J. 2017,
	\newblock {\em \apj} {\bf 849(2)}, 144
	
	\bibitem[\protect\astroncite{{Chen} and {Zhao}}{2018}]{Chen2018}
	{Chen}, R. and {Zhao}, J. 2018,
	\newblock {\em \apj} {\bf 853(2)}, 161
	
	\bibitem[\protect\astroncite{{Durney}}{1999}]{Durney1999}
	{Durney}, B.~R. 1999,
	\newblock {\em \apj} {\bf 511}, 945
	
	\bibitem[\protect\astroncite{{Duvall} et~al.}{1993}]{Duvall1993}
	{Duvall}, T.~L., J., {Jefferies}, S.~M., {Harvey}, J.~W., and {Pomerantz},
	M.~A. 1993,
	\newblock {\em \nat} {\bf 362(6419)}, 430
	
	\bibitem[\protect\astroncite{{Getling} et~al.}{2021}]{Getling2021}
	{Getling}, A.~V., {Kosovichev}, A.~G., and {Zhao}, J. 2021,
	\newblock {\em \apjl} {\bf 908(2)}, L50
	
	\bibitem[\protect\astroncite{{Ghizaru} et~al.}{2010}]{Ghizaru2010}
	{Ghizaru}, M., {Charbonneau}, P., and {Smolarkiewicz}, P.~K. 2010,
	\newblock {\em \apjl} {\bf 715(2)}, L133
	
	\bibitem[\protect\astroncite{{Gizon} et~al.}{2020a}]{Gizon2020a}
	{Gizon}, L., {Cameron}, R.~H., {Pourabdian}, M., {Liang}, Z.-C., {Fournier},
	D., {Birch}, A.~C., and {Hanson}, C.~S. 2020a,
	\newblock {\em Science} {\bf 368(6498)}, 1469
	
	\bibitem[\protect\astroncite{{Gizon} et~al.}{2020b}]{Gizon2020}
	{Gizon}, L., {Cameron}, R.~H., {Pourabdian}, M., {Liang}, Z.-C., {Fournier},
	D., {Birch}, A.~C., and {Hanson}, C.~S. 2020b,
	\newblock {\em Science} {\bf 368(6498)}, 1469
	
	\bibitem[\protect\astroncite{{Gough}}{1969}]{Gough1969}
	{Gough}, D.~O. 1969,
	\newblock {\em Journal of Atmospheric Sciences} {\bf 26(3)}, 448
	
	\bibitem[\protect\astroncite{{Gough} and {Toomre}}{1983}]{Gough1983}
	{Gough}, D.~O. and {Toomre}, J. 1983,
	\newblock {\em \solphys} {\bf 82(1-2)}, 401
	
	\bibitem[\protect\astroncite{{Guerrero} et~al.}{2016}]{Guerrero2016}
	{Guerrero}, G., {Smolarkiewicz}, P.~K., {de Gouveia Dal Pino}, E.~M.,
	{Kosovichev}, A.~G., and {Mansour}, N.~N. 2016,
	\newblock {\em \apj} {\bf 819(2)}, 104
	
	\bibitem[\protect\astroncite{{Guerrero} et~al.}{2013}]{Guerrero2013}
	{Guerrero}, G., {Smolarkiewicz}, P.~K., {Kosovichev}, A.~G., and {Mansour},
	N.~N. 2013,
	\newblock {\em \apj} {\bf 779(2)}, 176
	
	\bibitem[\protect\astroncite{{Guerrero} et~al.}{2019}]{Guerrero2019}
	{Guerrero}, G., {Zaire}, B., {Smolarkiewicz}, P.~K., {de Gouveia Dal Pino},
	E.~M., {Kosovichev}, A.~G., and {Mansour}, N.~N. 2019,
	\newblock {\em \apj} {\bf 880(1)}, 6
	
	\bibitem[\protect\astroncite{{Hartlep} et~al.}{2013}]{Hartlep2013}
	{Hartlep}, T., {Zhao}, J., {Kosovichev}, A.~G., and {Mansour}, N.~N. 2013,
	\newblock {\em \apj} {\bf 762(2)}, 132
	
	\bibitem[\protect\astroncite{{Hartlep} et~al.}{2008}]{Hartlep2008}
	{Hartlep}, T., {Zhao}, J., {Mansour}, N.~N., and {Kosovichev}, A.~G. 2008,
	\newblock {\em \apj} {\bf 689(2)}, 1373
	
	\bibitem[\protect\astroncite{{Hill}}{1989}]{Hill1989}
	{Hill}, F. 1989,
	\newblock {\em \apjl} {\bf 343}, L69
	
	\bibitem[\protect\astroncite{{Jackiewicz} et~al.}{2015}]{Jackiewicz2015a}
	{Jackiewicz}, J., {Serebryanskiy}, A., and {Kholikov}, S. 2015,
	\newblock {\em \apj} {\bf 805(2)}, 133
	
	\bibitem[\protect\astroncite{{Kholikov} et~al.}{2014}]{Kholikov2014b}
	{Kholikov}, S., {Serebryanskiy}, A., and {Jackiewicz}, J. 2014,
	\newblock {\em \apj} {\bf 784(2)}, 145
	
	\bibitem[\protect\astroncite{{Khomenko} et~al.}{2009}]{Khomenko2009}
	{Khomenko}, E., {Kosovichev}, A., {Collados}, M., {Parchevsky}, K., and
	{Olshevsky}, V. 2009,
	\newblock {\em \apj} {\bf 694(1)}, 411
	
	\bibitem[\protect\astroncite{{Kitchatinov} et~al.}{1994}]{Kitchatinov1994}
	{Kitchatinov}, L.~L., {Pipin}, V.~V., and {Ruediger}, G. 1994,
	\newblock {\em Astronomische Nachrichten} {\bf 315}, 157
	
	\bibitem[\protect\astroncite{{Kitiashvili} et~al.}{2015}]{Kitiashvili2015}
	{Kitiashvili}, I.~N., {Kosovichev}, A.~G., {Mansour}, N.~N., and {Wray}, A.~A.
	2015,
	\newblock {\em \apj} {\bf 809(1)}, 84
	
	\bibitem[\protect\astroncite{{Komm}}{2021}]{Komm2021}
	{Komm}, R. 2021,
	\newblock {\em \solphys} {\bf 296(12)}, 174
	
	\bibitem[\protect\astroncite{{Komm} et~al.}{2012}]{Komm2012}
	{Komm}, R., {Gonz{\'a}lez Hern{\'a}ndez}, I., {Hill}, F., {Bogart}, R.,
	{Rabello-Soares}, M.~C., and {Haber}, D. 2012,
	\newblock {\em \solphys} p. 177
	
	\bibitem[\protect\astroncite{{Kosovichev} and {Duvall}}{1997}]{Kosovichev1997}
	{Kosovichev}, A.~G. and {Duvall}, T.~L., J. 1997,
	\newblock in F.~P. {Pijpers}, J. {Christensen-Dalsgaard}, and C.~S. {Rosenthal}
	(eds.), {\em SCORe'96 : Solar Convection and Oscillations and their
		Relationship}, Vol. 225 of {\em Astrophysics and Space Science Library}, pp
	241--260
	
	\bibitem[\protect\astroncite{{Kosovichev} and {Pipin}}{2019}]{Kosovichev2019}
	{Kosovichev}, A.~G. and {Pipin}, V.~V. 2019,
	\newblock {\em \apjl} {\bf 871(2)}, L20
	
	\bibitem[\protect\astroncite{{Kosovichev} and {Zhao}}{2016}]{Kosovichev2016}
	{Kosovichev}, A.~G. and {Zhao}, J. 2016,
	\newblock in J.-P. {Rozelot} and C. {Neiner} (eds.), {\em Lecture Notes in
		Physics, Berlin Springer Verlag}, Vol. 914, p.~25
	
	\bibitem[\protect\astroncite{Krause and R\"adler}{1980}]{Krause1980}
	Krause, F. and R\"adler, K.-H. 1980,
	\newblock {\em Mean-Field Magnetohydrodynamics and Dynamo Theory},
	\newblock Berlin: Akademie-Verlag
	
	\bibitem[\protect\astroncite{{Lin} and {Chou}}{2018}]{Lin2018a}
	{Lin}, C.-H. and {Chou}, D.-Y. 2018,
	\newblock {\em \apj} {\bf 860(1)}, 48
	
	\bibitem[\protect\astroncite{{Miesch} et~al.}{2011}]{Miesch2011}
	{Miesch}, M.~S., {Brown}, B.~P., {Browning}, M.~K., {Brun}, A.~S., and
	{Toomre}, J. 2011,
	\newblock in N.~H. {Brummell}, A.~S. {Brun}, M.~S. {Miesch}, and Y. {Ponty}
	(eds.), {\em IAU Symposium}, Vol. 271 of {\em IAU Symposium}, pp 261--269
	
	\bibitem[\protect\astroncite{{Parchevsky} and
		{Kosovichev}}{2009}]{Parchevsky2009}
	{Parchevsky}, K.~V. and {Kosovichev}, A.~G. 2009,
	\newblock {\em \apj} {\bf 694(1)}, 573
	
	\bibitem[\protect\astroncite{{Parchevsky} et~al.}{2014}]{Parchevsky2014}
	{Parchevsky}, K.~V., {Zhao}, J., {Hartlep}, T., and {Kosovichev}, A.~G. 2014,
	\newblock {\em \apj} {\bf 785(1)}, 40
	
	\bibitem[\protect\astroncite{{Parker}}{1955}]{Parker1955}
	{Parker}, E.~N. 1955,
	\newblock {\em \apj} {\bf 122}, 293
	
	\bibitem[\protect\astroncite{{Pipin}}{2004}]{Pipin2004}
	{Pipin}, V.~V. 2004,
	\newblock {\em Astronomy Reports} {\bf 48}, 418
	
	\bibitem[\protect\astroncite{{Pipin}}{2021}]{Pipin2021}
	{Pipin}, V.~V. 2021,
	\newblock {\em arXiv e-prints} p. arXiv:2112.09460
	
	\bibitem[\protect\astroncite{{Pipin} and {Kitchatinov}}{2000}]{Pipin2000}
	{Pipin}, V.~V. and {Kitchatinov}, L.~L. 2000,
	\newblock {\em Astronomy Reports} {\bf 44}, 771
	
	\bibitem[\protect\astroncite{{Pipin} and {Kosovichev}}{2011}]{Pipin2011}
	{Pipin}, V.~V. and {Kosovichev}, A.~G. 2011,
	\newblock {\em \apjl} {\bf 727(2)}, L45
	
	\bibitem[\protect\astroncite{{Pipin} and {Kosovichev}}{2018}]{Pipin2018}
	{Pipin}, V.~V. and {Kosovichev}, A.~G. 2018,
	\newblock {\em \apj} {\bf 854(1)}, 67
	
	\bibitem[\protect\astroncite{{Pipin} and {Kosovichev}}{2019}]{Pipin2019}
	{Pipin}, V.~V. and {Kosovichev}, A.~G. 2019,
	\newblock {\em \apj} {\bf 887(2)}, 215
	
	\bibitem[\protect\astroncite{{Pipin} and {Kosovichev}}{2020}]{Pipin2020}
	{Pipin}, V.~V. and {Kosovichev}, A.~G. 2020,
	\newblock {\em \apj} {\bf 900(1)}, 26
	
	\bibitem[\protect\astroncite{{Reinecke} and {Seljebotn}}{2013}]{Reinecke2013}
	{Reinecke}, M. and {Seljebotn}, D.~S. 2013,
	\newblock {\em \aap} {\bf 554}, A112
	
	\bibitem[\protect\astroncite{{Rempel}}{2006}]{Rempel2006}
	{Rempel}, M. 2006,
	\newblock {\em \apj} {\bf 647}, 662
	
	\bibitem[\protect\astroncite{{Rempel} et~al.}{2009}]{Rempel2009}
	{Rempel}, M., {Sch{\"u}ssler}, M., and {Kn{\"o}lker}, M. 2009,
	\newblock {\em \apj} {\bf 691(1)}, 640
	
	\bibitem[\protect\astroncite{{Russell} et~al.}{2020}]{Russell2020}
	{Russell}, C.~T., {Luhmann}, J.~G., and {Jian}, L.~K. 2020,
	\newblock in A. {Kosovichev}, S. {Strassmeier}, and M. {Jardine} (eds.), {\em
		Solar and Stellar Magnetic Fields: Origins and Manifestations}, Vol. 354, pp
	127--133
	
	\bibitem[\protect\astroncite{{Schad} et~al.}{2013}]{Schad2013a}
	{Schad}, A., {Timmer}, J., and {Roth}, M. 2013,
	\newblock {\em \apjl} {\bf 778(2)}, L38
	
	\bibitem[\protect\astroncite{{Schatten} et~al.}{1978}]{Schatten1978}
	{Schatten}, K.~H., {Scherrer}, P.~H., {Svalgaard}, L., and {Wilcox}, J.~M.
	1978,
	\newblock {\em \grl} {\bf 5(5)}, 411
	
	\bibitem[\protect\astroncite{{Schou} et~al.}{1998}]{Schou1998}
	{Schou}, J., {Antia}, H.~M., {Basu}, S., {Bogart}, R.~S., {Bush}, R.~I.,
	{Chitre}, S.~M., {Christensen-Dalsgaard}, J., {Di Mauro}, M.~P.,
	{Dziembowski}, W.~A., {Eff-Darwich}, A., {Gough}, D.~O., {Haber}, D.~A.,
	{Hoeksema}, J.~T., {Howe}, R., {Korzennik}, S.~G., {Kosovichev}, A.~G.,
	{Larsen}, R.~M., {Pijpers}, F.~P., {Scherrer}, P.~H., {Sekii}, T., {Tarbell},
	T.~D., {Title}, A.~M., {Thompson}, M.~J., and {Toomre}, J. 1998,
	\newblock {\em \apj} {\bf 505(1)}, 390
	
	\bibitem[\protect\astroncite{{Simitev} et~al.}{2015}]{Simitev2015}
	{Simitev}, R.~D., {Kosovichev}, A.~G., and {Busse}, F.~H. 2015,
	\newblock {\em \apj} {\bf 810(1)}, 80
	
	\bibitem[\protect\astroncite{{Smolarkiewicz}}{2006}]{Smolarkiewicz2006}
	{Smolarkiewicz}, P.~K. 2006,
	\newblock {\em International Journal for Numerical Methods in Fluids} {\bf
		50(10)}, 1123
	
	\bibitem[\protect\astroncite{{Smolarkiewicz} and
		{Charbonneau}}{2013}]{Smolarkiewicz2013}
	{Smolarkiewicz}, P.~K. and {Charbonneau}, P. 2013,
	\newblock {\em Journal of Computational Physics} {\bf 236}, 608
	
	\bibitem[\protect\astroncite{{Stein} and {Nordlund}}{2012}]{Stein2012}
	{Stein}, R.~F. and {Nordlund}, {\AA}. 2012,
	\newblock {\em \apjl} {\bf 753}, L13
	
	\bibitem[\protect\astroncite{{Stejko} et~al.}{2021a}]{Stejko2021}
	{Stejko}, A.~M., {Kosovichev}, A.~G., and {Mansour}, N.~N. 2021a,
	\newblock {\em \apjs} {\bf 253(1)}, 9
	
	\bibitem[\protect\astroncite{{Stejko} et~al.}{2021b}]{Stejko2021a}
	{Stejko}, A.~M., {Kosovichev}, A.~G., and {Pipin}, V.~V. 2021b,
	\newblock {\em \apj} {\bf 911(2)}, 90
	
	\bibitem[\protect\astroncite{{Stenflo}}{2012}]{Stenflo2012}
	{Stenflo}, J.~O. 2012,
	\newblock {\em \aap} {\bf 547}, A93
	
	\bibitem[\protect\astroncite{{Zhao} et~al.}{2013}]{Zhao2013}
	{Zhao}, J., {Bogart}, R.~S., {Kosovichev}, A.~G., {Duvall}, T.~L., J., and
	{Hartlep}, T. 2013,
	\newblock {\em \apjl} {\bf 774(2)}, L29
	
	\bibitem[\protect\astroncite{{Zhao} et~al.}{2014}]{Zhao2014}
	{Zhao}, J., {Kosovichev}, A.~G., and {Bogart}, R.~S. 2014,
	\newblock {\em \apjl} {\bf 789(1)}, L7
	
\end{thebibliography}

\end{document}